\documentclass[aps,twocolumn,preprintnumbers,groupedaddress,amsmath,amssymb]{revtex4-2}
\usepackage{graphicx}
\usepackage[svgnames]{xcolor}
\usepackage{bm}
\usepackage{multirow}
\usepackage{hyperref}
\usepackage{pdfpages}
\usepackage{lineno}

\newif\ifshowcomments\showcommentstrue

\makeatletter
\AtBeginDocument{\let\LS@rot\@undefined}
\makeatother

\begin{document}

\title{Field-Induced Lifshitz Transition in the Magnetic Weyl Semimetal Candidate PrAlSi}

\author{Lei Wu$^{1}$, \textcolor{black}{Shengwei Chi$^{1}$,} Huakun Zuo$^{1}$,\textcolor{black}{Gang Xu$^{1}$}, Lingxiao Zhao\textcolor{black}{$^{2,*}$}, Yongkang Luo$^{1,*}$, Zengwei Zhu$^{1,*}$ }

	\affiliation{(1) Wuhan National High Magnetic Field Center and School of Physics, Huazhong University of Science and Technology,Wuhan 430074, China\\
		(2) Department of Physics,Southern University of Science and Technology, Shenzhen 518055, China\\
	}

\date{\today}
\begin{abstract}

Lifshitz transition (LT) refers to an abrupt change in the electronic structure and Fermi surface, and is associated to a variety of emergent quantum phenomena. Amongst the LTs observed in known materials, the field-induced LT has been rare and its origin remains elusive. To understand the origin of field-induced LT, it is important to extend the material basis beyond the usual setting of heavy fermion metals. Here, we report on a field-induced LT in PrAlSi, a magnetic Weyl semimetal candidate with localized 4$f$ electrons, through a study of magnetotransport up to 55 T. The quantum oscillation analysis reveals that across a threshold field $B^*\approx$14.5 T the oscillation frequency ($F_1$ = 43 T) is replaced by two new frequencies ($F_2$ = 62 T and $F_3$ = 103 T). Strikingly, the LT occurs well below quantum limit, with obvious temperature-dependent oscillation frequency and field-dependent cyclotron mass. Our work not only enriches the rare examples of field-induced LTs, but also paves the way for further investigation on the interplay among topology, magnetism and electronic correlation.
\end{abstract}

\maketitle
The Lifshitz transition (LT) has received renewed attention in the condensed matter physics. A LT \cite{Lifshitz1960} is an electronic topological transition of the Fermi surface (FS) driven by the variation of the band structure and/or the Fermi energy. Since such a transition does not necessarily require simultaneous symmetry breaking, and meanwhile, it can occur at $T=0$, be tuned by parameters other than temperature (such as pressure, strain, doping, magnetic field etc.)\cite{Lifshitz1960,Varlamov2021}, it, therefore, can be deemed as a topological quantum phase transition. In the vicinity of Lifshitz transitions, many peculiar emergent phenomena may appear, such as van-Hove singularity, non-Fermi-liquid behavior, unconventional superconductivity and so on (e.g. \cite{Steppke2017,LuoY-Sr2RuO4NMR}).


Compared with a number of cases tuned by doping or pressure that have been widely seen in topological systems~\cite{Yang2019,Liu2020}, cuprate superconductors~\cite{Benhabib2015,Norman2010}, iron pnictides superconductors~\cite{Ren2017,Liu2010}, and other strongly correlated materials~\cite{Steppke2017,Kwon2019}, the examples of LT driven by magnetic field are rare. This is because the energy scale of a laboratory magnetic field, in the order of 1-10 meV, is much smaller than the characteristic energy scale of most metals ($\sim 10^2-10^3$ meV). Only in a few cases, mostly limited in heavy-fermion (HF) metals~\cite{Kozlova2005,Aoki2016,Aoki1993,Bastien2016,Yelland2011,Niu2020,Pfau2013,Pfau2017,Gourgout2016}, the hybridization between conduction electrons and localized $f$ electrons leads to narrow renormalized bands with a small Fermi energy and thus the Zeeman term can be sufficiently strong to shift the spin-split FS \cite{Bastien2016}. Recently, field-induced LTs were also observed in some low Fermi energy non-magnetic semimetals such as bismuth \cite{Zhu2017}, TaP\cite{Zhang2017TaP} and TaAs\cite{Ramshaw2018}, wherein magnetic field beyond quantum limit can empty a Dirac or Weyl pocket with small Fermi energy. However, in these cases, no additional Fermi pocket emerges and the carriers of the empty pocket were transferred to other pockets (previously existing).

Here we present a new example of field-induced LT beyond heavy-fermion systems in the magnetic Weyl semimetal candidate PrAlSi, by a systematic study of quantum oscillation (QO) effect with the magnetic field extending up to 55 T. We observe a single frequency ($F_1$ = 43 T) below a critical field of $B^{\ast}=$14.5 T, in agreement with what was previously reported \cite{lyu2020}. Above $B^{\ast}$, we see clearly the emergence of two new frequencies ($F_2$ = 62 T and $F_3$ = 103 T)  and the disappearance of the original $F_1$. We exclude the possibility of magnetic breakdown and identify $B^{\ast}$ as a critical point where the field-induced LT occurs. By comparing the reported Fermi surface of NdAlSi and theoretical calculation of PrAlSi, we conclude that the LT occurs in the hole-like Weyl pockets along the direction of $\Gamma$-X of the Brillouin zone (BZ, hereafter). Our work not only enriches the rare examples of field-induced LTs, but also paves the way for further investigation on the interplay among topology, magnetism and electronic correlation.

The $R$Al$X$ family ($R$ = rare earth and $X$ = Si or Ge) compounds have recently been proposed to host ideal candidates of magnetic Weyl semimetals\cite{chang2018,puphal2019}, and provide a platform to investigate the interaction between magnetism and Weyl physics \cite{Nakatsuji2015,Kuroda2017,Liu2018,Morali2019,Sakai2018}. They crystallize in a tetragonal structure with the noncentrosymmetric space group symmetry of I4$_1$/md (No. 109). One advantage of this family is that Weyl nodes generated by inversion breaking are robust and can be shifted by the Zeeman coupling in the $k$ space~\cite{chang2018}. Several intriguing physical properties have been observed in this family. The list includes the coexistence of type-\uppercase\expandafter{\romannumeral1} and type-\uppercase\expandafter{\romannumeral2} Weyl fermion in LaAlSi~\cite{su2021} and LaAlGe~\cite{Xu2021LaAlGe}, topological Hall effect in CeAlGe\cite{Puphal-CeAlGeTHE}, anisotropic anomalous Hall effect in CeAlSi\cite{Yang2021}, and Weyl-driven collective magnetism in NdAlSi~\cite{gaudet2020}. The compound PrAlSi studied here is a ferromagnetic semimetal with Curie temperature $T_{\rm{C}}\sim$18 K. A recent work based on static field (9 T) transport measurements by Lyu et al. revealed a large anomalous Hall conductivity $\sim$2000 $\Omega^{-1}{\rm cm^{-1}}$ and an unusual temperature dependence of QO with a single frequency~\cite{lyu2020}.

\begin{figure}[!t]
	
\includegraphics[width=8.5cm]{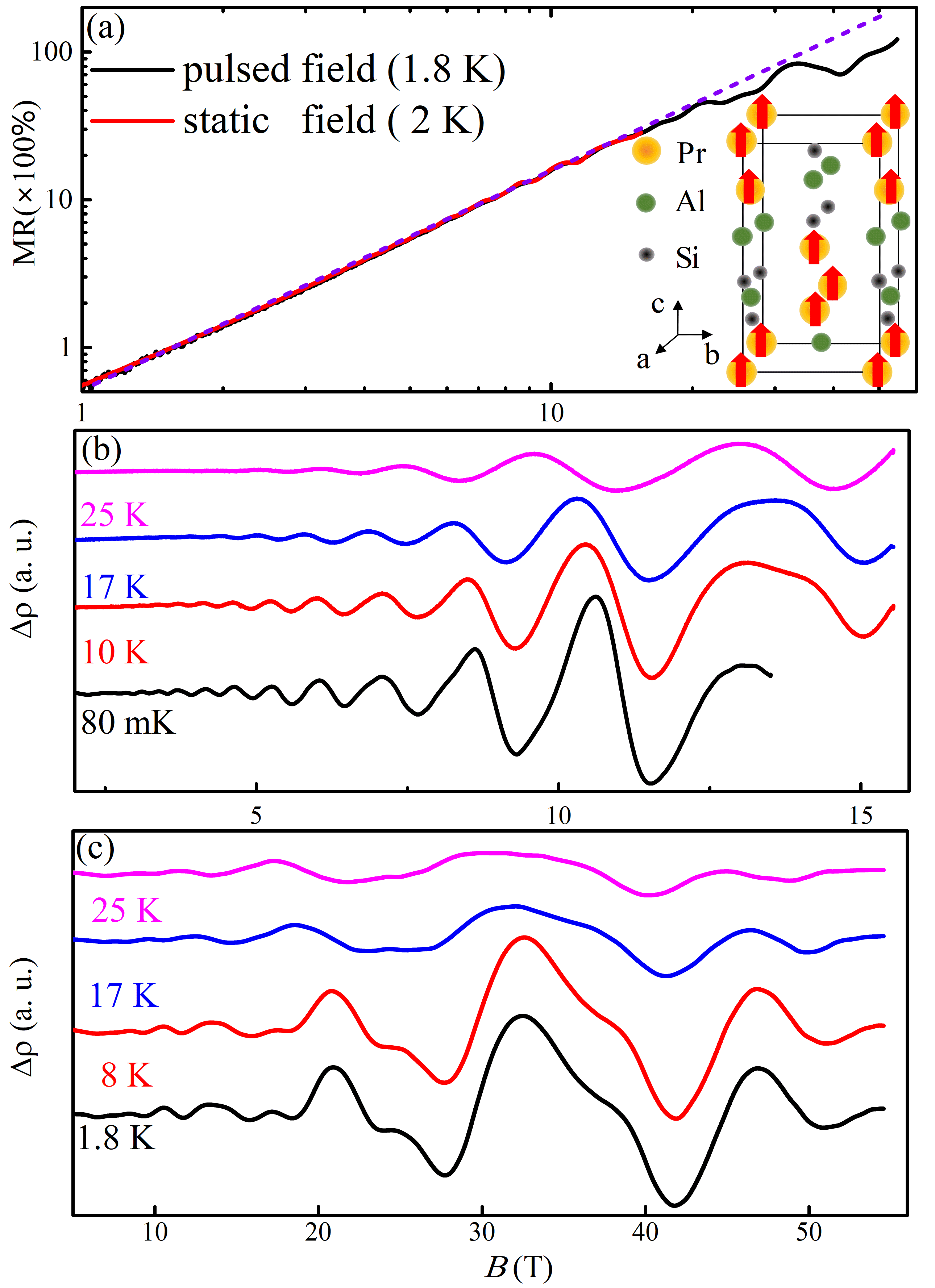}

\caption{(a) The field-dependent of MR measured under static field at 2 K and pulsed field at 1.8 K. The inset is crystal structure(I4$_1$/md) of PrAlSi as the magnetic field easily polarizes the moments of Pr along the $c$-axis. (b) Field dependence of the oscillatory part of magnetoresistance $\Delta\rho(B)$ in low magnetic field measured under static field, showing only a single QO frequency. The experiments were carried out with a Leiden dilution refrigerator (14 T) at 80 mK and with a refrigerator of Oxford Instrument (16 T) at $T$ = 10 K, 17 K and 25 K. (c) $\Delta\rho(B)$ in high magnetic field measured by pulsed magnetic field, showing more QO frequencies with a complex pattern.
\label{fig:longitude}}
\end{figure}

High-quality single crystals of PrAlSi were synthesized using the flux method. The inset of Fig. \ref{fig:longitude}(a) shows the crystal structure of PrAlSi with the magnetic moments of Pr easily orientated along $c$-axis after the application of a small field \cite{lyu2020}. In our transport measurements, the magnetic field was along the $c$-axis and the electrical current was along the $b$-axis (more details are presented in METHODS and Supplemental Material). Fig. \ref{fig:longitude}(a) shows magnetoresistance measured in static field at 2 K (red curve) and pulsed field at 1.8 K (black curve). Normalized magnetoresistance $(\rho(B)-\rho(0))/\rho(0)$ reaches 116 at 55 T and remains non-saturating. The purple dashed line corresponds to $B^{1.7}$. Fig. \ref{fig:longitude}(b) presents the oscillatory part of the longitudinal resistivity $\Delta\rho(B)$, obtained by subtracting a smooth background from the measured $\rho(B)$ (see the Supplementary Figure 2 for more raw data). At low temperature, oscillations are visible above the field as low as 3 T, indicating the good quality of the sample. The rough Dingle mobility of 1/$B\rm{_c}$ = 0.33 T$^{-1}$ is close to the average mobility of 0.26 m$^2$/Vs yielded from the amplitude of the quadratic low-field magnetoresistance (see the inset of the Supplementary Figure 2 (a)). The single-frequency QO for static field measurements retains until temperature down to 80 mK. A more complex pattern emerges when larger magnetic field is applied, as seen in Fig. \ref{fig:longitude}(c).

\begin{figure}
\includegraphics[width=8.5cm]{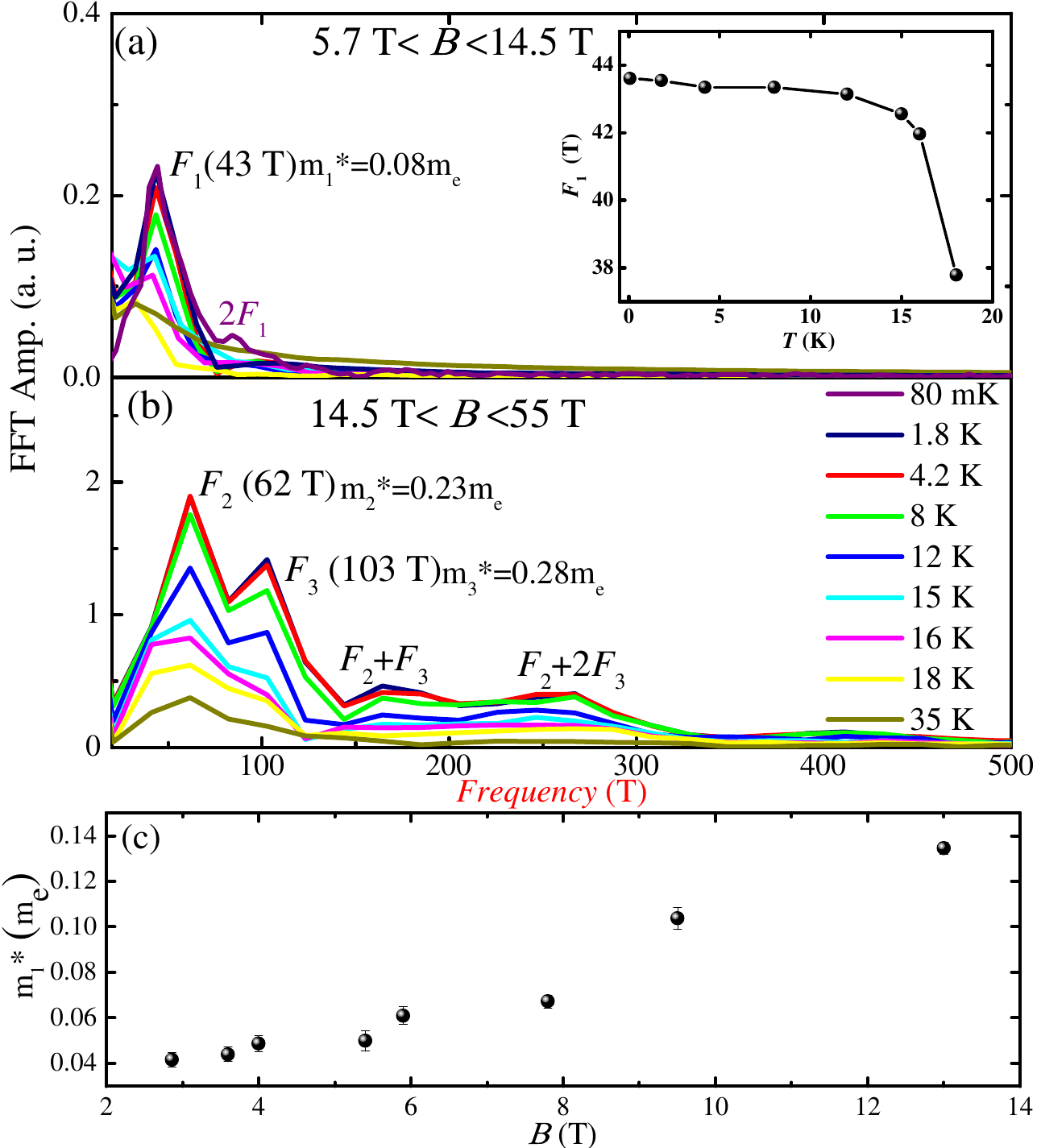}
\caption {(a)(b) The frequency (T) dependence of the fast-Fourier-transformation amplitude of SdH oscillations measured at various temperatures with $B$ $\parallel$ $c$ in pulsed field. With the field lower than 14.5 T, only one fundamental frequency exists in (a). Two main higher frequencies and harmonic terms are shown in (b) with the field above 14.5 T. The inset shows the temperature-dependent frequency $F_1$ extracted from (a). (c) The corresponding effective mass of $F_1$ as a function of magnetic field $B$ obtained from the analysis of the QO peaks in static field.}
\label{Fig:FFT}
\end{figure}
Fig. \ref{Fig:FFT}(a) and (b) show the results of the fast Fourier transformation (FFT) of the oscillatory part of the magnetoresistance $\Delta\rho$ as a function of $1/B$. The SdH frequencies extracted from low-field ($B <$ 14.5 T) and high-field ($B>$ 14.5 T) data display a dramatic difference. Note that, since this compound is ferromagnetic, we took into the demagnetization factor to correct the applied field in all the analyses of SdH effect (see the Section 9 of the Supplemental Material). We identified $B^{\ast}=$14.5 T as a critical field, after checking several fields close to $B^{\ast}$. As shown in the Supplementary Figure 3, we have intercepted a field with different values, 10, 14, 14.5, 15 and 20 T for the segmented FFT analyses. We can see the crossover from the lower QO frequency to the two higher QO frequencies as the segment changes between the 5.7 T and the selected field. Two higher QO peaks emerge when we segment the field at 20 T and we note that the low QO peaks persist because of the inclusion of the low-field QOs below 14.5 T. From the comparison of the segmented FFT analyses, ~14.5 T was determined as the field of the LT. There is only one fundamental frequency in the FFT spectrum until down to 80 mK below $B^{\ast}$, see Fig. \ref{Fig:FFT}(a). It should be noted that $F_1$ gradually decreases with temperature, changing from 43 T at 80 mK to 32 T at 35 K, as shown in the inset. Similar temperature dependence in $F_1$ was also reported in an earlier work on PrAlSi\cite{lyu2020}, whereas the values of $F_1$ are relatively smaller than ours. We attribute this discrepancy to the difference in stoichiometry\cite{Pfau2017}.

Fig. \ref{Fig:FFT}(b) shows that above $B^{\ast}$ there are two new QO frequencies ($F_2$ = 62 T, $F_3$ = 103 T) and their high-order harmonics. Such a change in Fermi surface is also manifested in the effective cyclotron mass $m^*$. The value of $m^*$ for each frequency can be deduced from the fitting of FFT amplitude according to the temperature damping factor, and this yields the small $m^*$=0.08 $m_e$ for $F_1$ (with field range 5.7-14.5 T), and 0.23 $m_e$ and 0.28 $m_e$ for $F_2$ and $F_3$, respectively, where $m_e$ is the mass of a free electron. Interestingly, a careful look into the temperature dependence of the amplitude of the oscillatory peak leads to the fact that the effective mass is enhanced 2-fold between 2.8 T and $B^{\ast}$, as shown in Fig. \ref{Fig:FFT}(c). This feature is reminiscent of HF systems displaying a LT, and will be discussed more later on. The effective masses of $F_2$ and $F_3$ as the function of field higher than $B^{\ast}$ are absent here, mainly because it's rather difficult to extract the exact amplitude of entangled peaks in oscillatory part from two frequencies and their harmonic terms. More explanation about effective mass mentioned above is exhibited in Supplementary Figure 5. The Fermi energy of the band corresponding to $F_1$ is then estimated $\varepsilon_F$$\sim$125 meV.

To further demonstrate the field-induced Fermi surface change near $B^*$=14.5 T, we performed Lifshitz-Kosevich (LK) fitting on $\Delta\rho(B)$ \textcolor{black}{measured under pulsed field. We notice that all the analyzed FFT data are from the pulsed fields.} More details can be found in Section 4 of the Supplemental Material. As is shown in Fig. \ref{fig:fitting}(a), the $\Delta\rho$ for field below $B^*$ can be well reproduced by the LK fitting with a single $F_1$ (cf. the red dot line). However, such a fitting collapses when field exceeds $B^*$. This problem can be fixed in an alternate fitting by employing both $F_2$ and $F_3$, seeing the blue dot line in Fig. \ref{fig:fitting}(b). Noteworthy that this 2-frequency LK fitting fails in the low-field window, implying that $F_2$ and $F_3$ appear only in the high-field range.

\begin{figure}
\includegraphics[width=8.5cm]{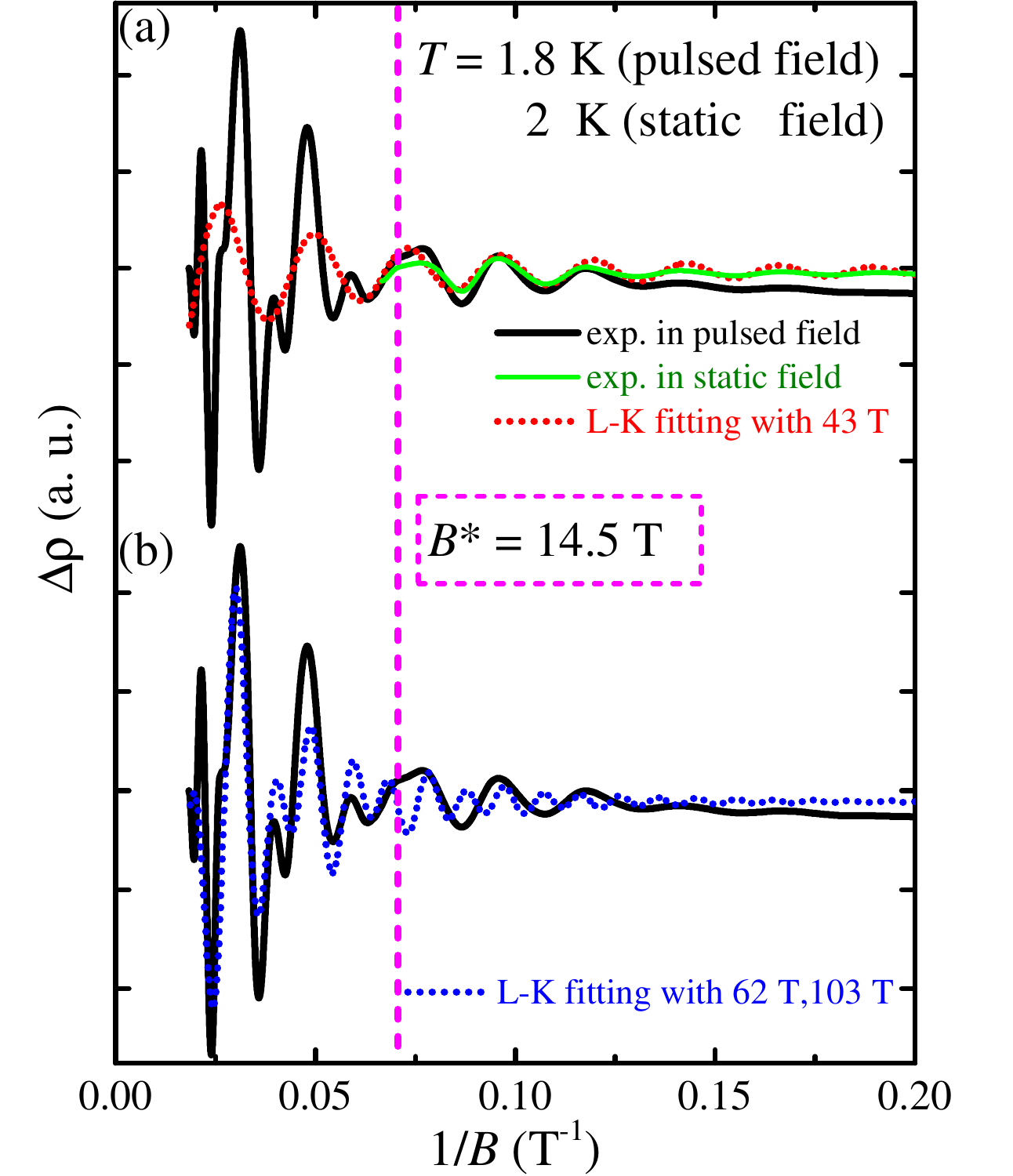}
\caption{The Lifshitz-Kosevich fit to $\Delta\rho(B)$ measured with pulsed field. Inverse field dependence of the oscillatory magnetoresistance measured at 1.8 K with pulsed field and at 2 K with static field are depicted as black and green lines, respectively. Note that the L-K fitting with a single frequency of 43 T reproduces the SdH oscillations nicely for field below $B^*$, but fails for $B>B^*$, as shown in panel (a). A multi-frequency L-K fitting by employing both $F_2$ and $F_3$ describes $\Delta\rho(B)$ reasonably well for $B>B^*$, seeing panel (b).}
\label{fig:fitting}
\end{figure}

\begin{figure}
\includegraphics[width=8.5cm]{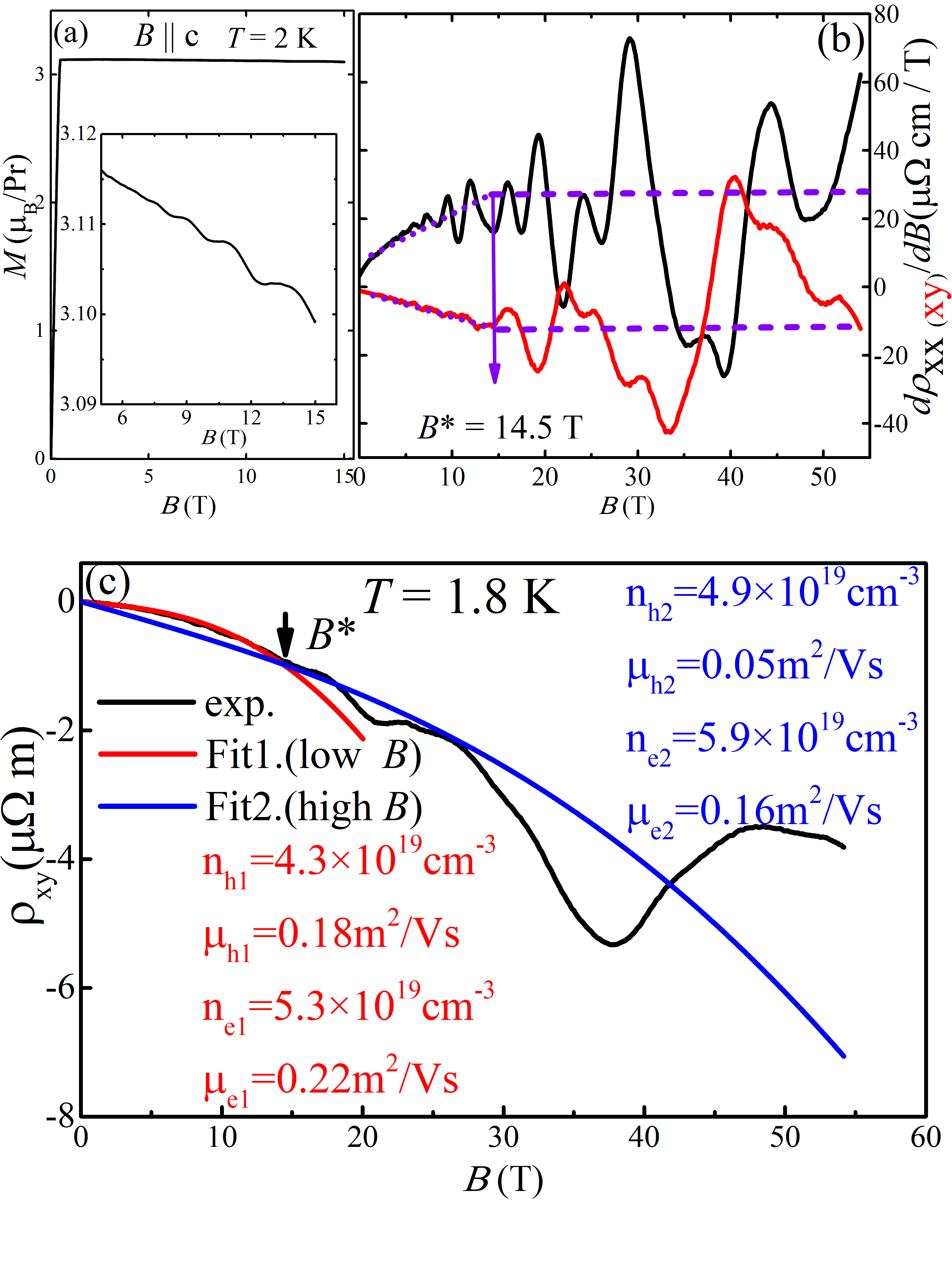}
\caption{(a) The magnetization of PrAlSi with the magnetic field applied along the $c$-axis, the inset is the enlarged part with evident QOs. (b) The black and red lines are the first derivative of the MR and Hall resistivities measured under pulsed field which could provide more information beyond the critical field. The violet dotted lines guided by eyes suggest the change of slopes in both of them at the $B^{\ast}$. (c) The field dependence of Hall resistance at 1.8 K along with two fitting curves with two-band model. The red and blue curves correspond to the fits to the low-field and high-field respectively.
\label{fig:scenario}}
\end{figure}

Thus, the variation of QO frequencies with the disappearance of 43 T and the emergence of 62 T and 103 T clearly point to the change in Fermi surface topology, viz a LT. Firstly, we can rule out magnetic breakdown as the origin, because there is no extra QO frequency (19 T and 60 T) in the low-field range even for temperature as low as 80 mK. Moreover, the peak with frequency of 43 T does not persist under high magnetic field, either.
Secondly, we can also exclude a metamagnetic transition as the driver of this process. Fig. \ref{fig:scenario}(a) shows the field dependence of magnetization at 2 K with the magnetic field applied along $c$-axis. One clearly finds that the magnetization saturates to $\sim$ 3.1 $\mu_B$/Pr at a small field 0.48 T, and no additional transition can be resolved nearby 14.5 T except for some traces of de Haas-van Alphen oscillations (inset of Fig. \ref{fig:scenario}(a)). This is different from the case of NdAlSi, where a magnetic transition to the final Weyl-mediated helical magnetism leads to a change in QO frequency~\cite{gaudet2020}.

In order to further clarify this field-induced LT, it is helpful to estimate the density of carriers of each Fermi pocket. According to the Lifshitz-Onsager relation, $F=(\hbar/2\pi{e})A_F$, where $\hbar$ is Planck's constant and $A_F=\pi k_F^2$ is an extremal cross-sectional area of the Fermi surface perpendicular to the field with Fermi wave vector $k_F$. The bands become non-degenerate due to spin-orbit coupling\cite{gaudet2020}. Assuming these Fermi pockets are spheres, we find that the LT wipes out $n_{F_1}$=3.2 $\times$10$^{18}$cm$^{-3}$ and produces $n_{F_2}$=5.5 $\times$10$^{18}$cm$^{-3}$ and $n_{F_3}$=1.2 $\times$10$^{19}$cm$^{-3}$ per 4 pockets. Note that the total number of pockets would be a multiple of four, due to the symmetric requirement (see below).

The \textit{total} carrier density of hole and electron can be also extracted by fitting the Hall resistivity to a two-band model,
$\rho{_{xy}}(B)=\frac{B}{\rm{e}}\frac{(n{_h}\mu{_h}^{2} - n{_e}\mu{_e}^{2})+\mu{_h}^{2}\mu{_e}^{2}(n{_h}-n{_e})B^2}{(n{_h}\mu{_h}+ n{_e}\mu{_e})^2+\mu{_h}^{2}\mu{_e}^{2}(n{_h}-n{_e})^2B^2}$. Here, $n$ and $\mu$ represent carrier density and mobility, and the subscripts $h$ and $e$ denote hole and electron, respectively. We obtain $n{_h}$= 4.3$\times$ 10$^{19}$ cm$^{-3}$, $n{_e}$= 5.3$\times$ 10$^{19}$ cm$^{-3}$, $\mu{_h}$= 0.18 m$^{2}$/Vs and $\mu{_e}$= 0.22 m$^{2}$/Vs. These values fit both Hall resistivity and magnetoresistivity reasonably well up to $B^{\ast}$ (see Fig. \ref{fig:scenario}(c) and Supplementary Figure 7). The deduced mobilities are also close to the value obtained from quantum oscillations and magnetoresistance. The average zero-field mobility $\langle\mu\rangle$ extracted from the residual resistivity $\rho_0$=15 $\mu\Omega\cdot$cm is about 0.5 m$^{2}$/Vs, slightly larger than the finite-field mobility. Such a discrepancy has been observed in other semimetals\cite{Ding2019,Fauque2018} and attributed to the field-induced mobility reduction. The carrier densities of hole and electron are within 10\% of the compensation $2[n_en_h/(n_e+n_h)]$, compared to$~\sim$ 4\% in bismuth\cite{Bhargava1967} and WTe$_2$\cite{Zhu2015}. This near compensation would explain the observed unsaturated magnetoresistance. The slight excess in hole may result from an uncontrollable doping which could be also the reason for the sample dependence of $F_1$ as discussed above.

This fit, which properly works up to $B^{\ast}$ (shown by a black arrow in Fig. \ref{fig:scenario}(c)), fails above $B^{\ast}$. The change occurring at $B^{\ast}$ is evident in Fig. \ref{fig:scenario}(b), which shows the first derivative of the longitudinal and Hall resistivities. The violet dotted lines demonstrate the apparent change of slope in MR and Hall resistivities at 14.5 T. We let $|n{_h}-n{_e}|$ to stay constant across $B^{\ast}$, in order to respect the Luttinger theorem \cite{Luttinger1960}. We infer that $F_2$ and $F_3$ should correspond to carriers of opposite signs with a density difference of $|n_{F_3}-n_{F_2}|=6.5\times$10$^{18}$ cm$^{-3}$. Assuming that there are 8 pockets for $F_1$ ($2 n_{F_1}$=6.4$\times$ 10$^{18}$cm$^{-3}$), would be compatible with the Luttinger theorem.  Both types of carriers increase by about 0.6$\times$10$^{19}$ cm$^{-3}$ ($n_{F_3}-2n_{F_1}$ and $n_{F_2}$, respectively). By fitting the Hall resistivity curve with $n{_h}$=4.9$\times$10$^{19}$cm$^{-3}$ and $n{_e}$ = 5.9$\times$10$^{19}$cm$^{-3}$, we obtain $\mu{_h}$=0.05(1) m$^{2}$/Vs and $\mu{_e}$= 0.16(1) m$^{2}$/Vs. This is shown in Fig. \ref{fig:scenario}(c) with blue line. The mobility ($\mu=\frac{e\tau}{m^{\ast}}$) of holes drop ($\sim$72\%) more than that of electron($\sim$27\%), implying the sign of $F_1$ with a small mass should be hole-like, since its mass increases by 2.5 times by assuming a same scattering time $\tau$. We conclude that $F_2$ has an electron-like sign and $F_3$ is a hole-like one. Thus, each hole pocket ($F_1$) evolves into a larger hole pocket ($F_3$) and an additional electron pocket ($F_2$). This indicates the existence of a van-Hove singularity (saddle point) in this system.

\begin{figure*}[ht]
\includegraphics[width=17cm]{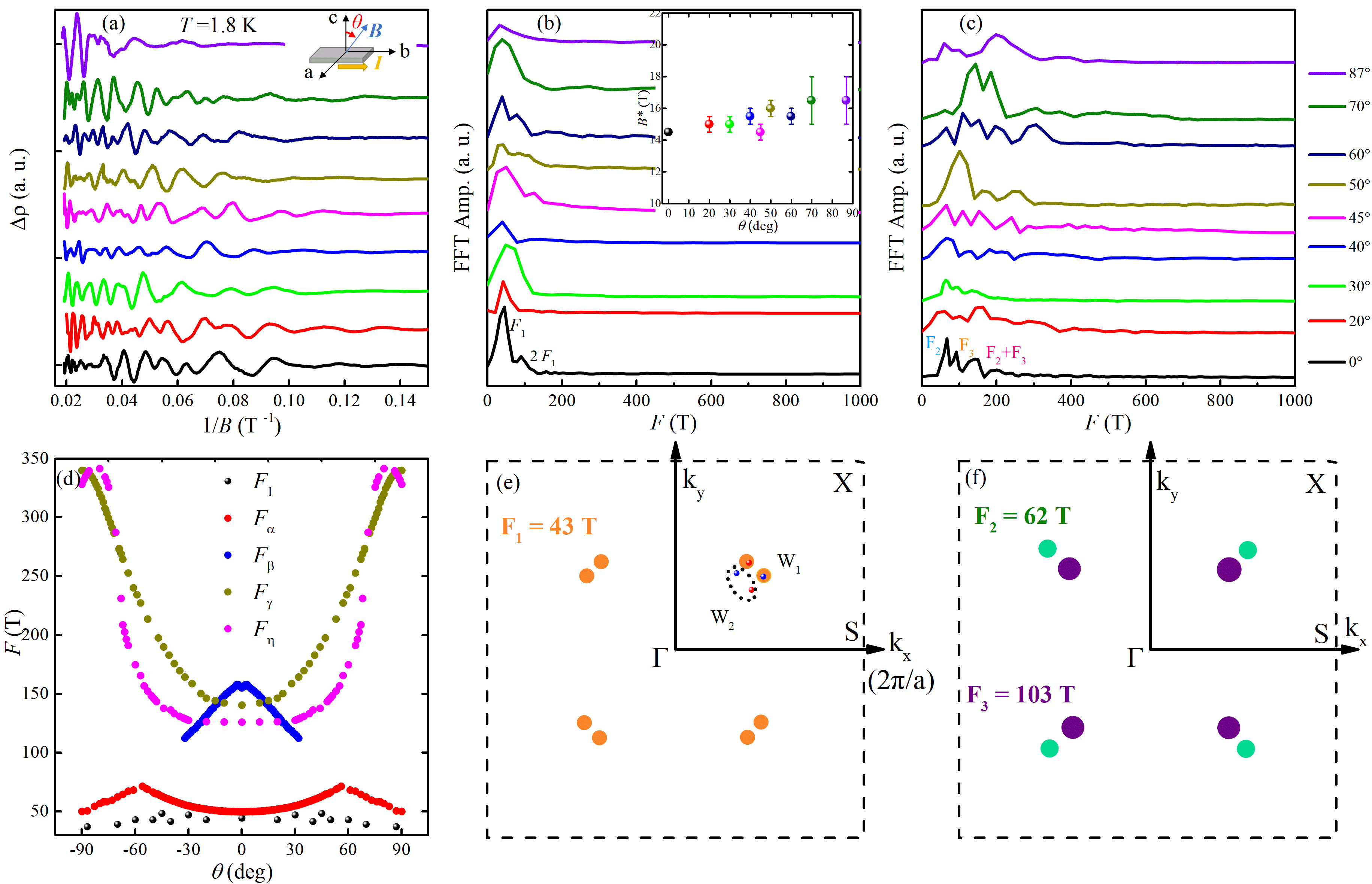}
\caption{(a) Background-subtracted SdH oscillatory part as a function of 1/$B$ at $T$ = 1.8 K under various angles. The inset on the top is the schematic diagram of the measurements. (b)(c) The FFT spectra for different angles and magnetic field ranges and the data are shifted along the vertical axis for clarity. The inset in graph (b) displays the critical fields of every angles and the error bars are defined as standard deviation. (b) The angular dependence of the corresponding measured oscillation frequency of $F_1$ pocket (black solid symbol) as well as the calculated frequencies $F_\alpha$, $F_\beta$, $F_\gamma$ and $F_\eta$.(e)(f) Schematic illustration of the Lifshitz transition in the BZ by the sketch of Fermi pockets which correspond to (b) and (c), respectively. Two tiny hole pockets (in orange) with $F_1=43$ T located along the sides of $\Gamma-$X transform into one electron pocket in green with $F_2=62$ T and one hole pocket in violet with $F_3=103$ T after LT, implying the existence of a van-Hove singularity. The pockets along  $\Gamma-$S are not shown here (see the Supplementary Figure 9). In graph (e), the projection of two pairs of Weyl points around the $\Gamma-$X line in the BZ are illustrated, in which W$_1$ and W$_2$ are indicated. Weyl points with opposite chiralities are marked as red and blue dots.}
\label{fig:angle}
\end{figure*}
To get more information about the Fermi surface of PrAlSi, we performed the measurements of angular-dependent MR with pulsed magnetic field at $T$= 1.8 K. The Fig. \ref{fig:angle} (a) shows the oscillatory component extracted by subtracting the smooth background from the MR measured at different {$\theta$}, which is defined as the angle between the $c$ axis and the magnetic field, and the current was along $b$ axis as shown in the inset (see the Supplementary Figure 10 for the raw MR data). The SdH oscillations evolve systematically and can be observed in all angles as the magnetic field is rotated from {$\theta$} = 0$^{\circ}$ to {$\theta$} = 87$^{\circ}$. Fig. \ref{fig:angle} (b) and Fig. \ref{fig:angle} (c) present the segmented FFT spectra for different angles {$\theta$}. The inset of figure (b) shows the critical fields at various angles, which indicating that the LT also envolves with $\theta$. We determined the $B^{\ast}$ as the same method mentioned above and the detail is shown in Supplementary Figure 4. The oscillation frequency $F_1$ shown in Fig. \ref{fig:angle} (b) is weakly dependent on angle, indeed indicating a nearly spherical Fermi surface of this band and the evolution of angle-dependent oscillation frequencies $F_2$, $F_3$ and their harmonic terms shown in Fig. \ref{fig:angle} (c). According to the experimental results, Fig. \ref{fig:angle} (d) presents the angle dependence of the quantum oscillation $F_1$ which is shown as black symbols as well as $F_\alpha$, $F_\beta$, $F_\gamma$ and $F_\eta$ are the calculated frequencies obtained from the SKEAF program \cite{julian2012numerical}. The Fermi level was shifted  with 3 meV in calculation because of the uncertain doping. The $F_\alpha$ is quite close to the observed $F_1$. Whereas the higher frequencies obtained from calculation are not been observed by experiment in our case, may be due to the low mobility of these bands.

Next, we tried to locate these pockets in the BZ, and our similar calculation results (see the Supplementary Figure 11). The electronic structure also resembles that of NdAlSi \cite{gaudet2020}. Since the center of the BZ is not occupied, symmetry imposes four-fold degeneracy of each pocket. According to the calculation, the schematic of the positions of pockets $F_1$, which are represented by orange spheres along the $\Gamma-$X direction of the BZ in Fig. \ref{fig:angle} (e). The pockets along $\Gamma-$S are large and cannot be easily modified, hence for clarity they are not shown here and the Fermi pockets are plotted in the Supplementary Figure 9. Instead, the bands along $\Gamma-$X are shallow and host Weyl points and they should be susceptible to magnetic field. The new $F_2$ and $F_3$ pockets appear after the disappearance of $F_1$. So it is a reasonable assumption that they locate on the $\Gamma-$X line as well where the pockets $F_1$ lie on, as shown in Fig. \ref{fig:angle} (f). Indeed, some Weyl signatures were observed for the pockets $F_1$. All the values extracted from the peak and valley positions of SdH oscillation fall on a line with an intercept of $\sim$-0.01 in the Landau fan diagram, indicating a non-trivial $\pi$ phase (see the Supplementary Figure 6(b)). Its cyclotron mass is very small and changes with magnetic field. Note that a normal parabolic band ($E=\frac{\hbar A_F}{2\pi m}$) would not change its cyclotron mass ($m_{CR}=\frac{\hbar^2}{2\pi}\frac{\partial A_F(E,k_{\parallel})}{\partial E}$)\cite{Singleton2001} as the Fermi energy changes. The increasing of the effective mass may imply the increase of spin fluctuation of the pockets with LT transition, similar to the early report \cite{Aoki1993}, instead of the response from a normal parabolic band. In order to better understand the Weyl nature of the Fermi pockets, it is indispensable for confirming the Weyl points along the direction of $\Gamma-$X line of the BZ. As shown in Fig. \ref{fig:angle} (e), we obtained two pairs of Weyl points along the $\Gamma-$X line by band structure calculation of selected paths of the BZ and the details are listed in the table of Supplementary Figure 11. The results agree with the prior work on Weyl points and Fermi surface of PrAlSi \cite{yang2020PrAlSi}.

Take together, three prominent features of the LT in PrAlSi can be found: (i) the QO frequencies are strongly temperature dependent, (ii) the quasiparticle effective mass increases gradually when approaching $B^{\ast}$, and (iii) the critical field for such LT is far below quantum limit. In conventional metals, only a small change in QO frequency of the order of ($k_B T/E\rm{_F}$)$^2$ is expected upon warming~\cite{shoenberg2009}.Temperature-dependent QO frequency was observed in some HF metals, which is typically ascribed to the sensitivity of the $f$-$c$ ($c$-conduction electrons) hybridization and the corresponding Kondo resonance state with a variation of temperature\cite{Goll2002,Kozlova2005,Aoki2016}. In this framework, the change of $m^*$ was also attributed to an itinerant - localized transition of $4f$ electrons\cite{Aoki1993}. However, in PrAlSi, DFT calculations manifested that the Pr-$4f$ bands locate well below the Fermi level (see the Supplementary Figure 8), notable $f$-$c$ hybridization is unlikely. Furthermore, PrAlSi is a low-carrier density semimetal, the Kondo screening is also expected to be weak according to Nozi\`{e}res exhaustion idea\cite{Nozieres-EPJB1998,He-PrBi}. Therefore, it is unlikely that the observed LT in PrAlSi originates from a competition between Zeeman term and a characteristic Kondo coherence energy as in HF systems. In the cases of bismuth \cite{Zhu2017}, TaP\cite{Zhang2017TaP} and TaAs\cite{Ramshaw2018}, the quantum limits have been reached to empty a Dirac or Weyl pocket, and caused a field-induced LT. In our case, however, a rough estimate yields the quantum limit field in the range of 40-100 T, much larger than the critical field $B^*\sim$14.5 T.

Another possibility for the observed LT in PrAlSi might be related to the crystal-electric-field (CEF) effect. This arises because the  nine-fold-degenerate $j$ = 4 multiplet of Pr$^{3+}$ in a $D_{2d}$ ($\bar{4}$2m) point symmetry CEF splits into two non-Kramers doublets and five singlets\cite{Dhar-PrSi2}. A recent analysis based on specific heat measurements revealed that the ground state is probably a doublet, while the magnetic entropy gain reaches $R\ln3$ at about 20 K, $R\ln4$ at about 30 K, and saturates to $R\ln9$ at a temperature as low as $\sim$95 K\cite{lyu2020}. This suggests that at least one excited state sitting not far above the ground doublet, potentially in the order of 10 K. It is reasonable to speculate that magnetic field of $\sim$ 10 T might be sufficient to modify the CEF energy levels and the orbital characters, which possibly changes the Fermi surface topology. In addition, this field-induced evolution of CEF levels is also qualitatively consistent with the temperature-dependent QO frequency and field dependent $m^*$ as observed experimentally. Actually, this scenario was also proposed recently for the field-induced Fermi surface reconstruction in CeRhIn$_5$\cite{Lesseux-CeRhIn5NMR}. To further address this possibility, more experiments like inelastic neutron scattering are needed to figure out the diagram of the CEF splitting.

In summary, we grew high quality single crystals and observed pronounced SdH oscillations in PrAlSi with magnetic field up to 55 T. A LT transition occurs around 14.5 T. The change in carrier densities and Fermi pockets revealed by QO and Hall effect are consistent with each other. By comparison with theoretical calculations, we propose that LT occurs along the $\Gamma$-X orientation and involves the Weyl pockets. One hole pocket becomes an electron pocket and a hole pocket, which indicates the existence of a van-Hove singularity. PrAlSi, therefore, represents a unique case of field-induced LT beyond the HF systems.

\noindent
 *\verb|zhaolx@mail.sustech.edu.cn|\\
 *\verb|mpzslyk@hust.edu.cn|\\
 *\verb|zengwei.zhu@hust.edu.cn|\\
\textbf{DATA AVAILABILITY}\\
The data supporting the present work are available from the corresponding authors upon request.\\

\textbf{ACKNOWLEDGEMENTS}\\
We thank Kamran Behina for insightful discussions. This work is supported by the National Key Research and Development Program of China (Grant No.2022YFA1403503),  the National Science Foundation of China (Grant Nos.12004123, 51861135104 and 11574097), the Fundamental Research Funds for the Central Universities (Grant no. 2019kfyXMBZ071), the open research fund of Songshan Lake Materials Laboratory (2022SLABFN27) and National Key R\&D Program of China (2022YFA1602602).\\

\textbf{COMPETING INTERESTS}\\
The authors declare no competing interests.\\

\textbf{AUTHOR CONTRIBUTIONS}\\
Z.Z. and L.Z. conceived and oversaw this work; L.W., H.Z. and L.Z. performed the experiments. Z.Z., \textcolor{black}{G.X, L.W. and S.C} performed band-structure calculations. L.W, L.Z., Y.L. and Z.Z. wrote the manuscript with inputs from co-authors.\\


\clearpage

\renewcommand{\thesection}{S\arabic{section}}
\renewcommand{\thetable}{S\arabic{table}}
\renewcommand{\thefigure}{S\arabic{figure}}
\renewcommand{\theequation}{S\arabic{equation}}

\setcounter{section}{0}
\setcounter{figure}{0}
\setcounter{table}{0}
\setcounter{equation}{0}

{\large\bf Supplementary information for ``Field-Induced Lifshitz Transition in the Magnetic Weyl Semimetal Candidate PrAlSi"}

\section{Samples and Methods}
Single crystals of PrAlSi used in our studies were synthesized using the flux method. The starting materials are high purity chunks of praseodymium, silicon and aluminum, mixed into an alumina crucible. Then, the alumina crucible and quartz wool were placed in a quartz tube, which was sealed under high vacuum, heated to 1100$^\circ$C at 3$^{\circ}$C/min. And holding for 12 h and then the tube was cooled down to 750$^\circ$C in 100 h and dwells for 2 days. The excess Al flux was removed by centrifuging. To identify the as-grown sample quality of PrAlSi, powder crystal x-ray diffraction (XRD) was used. And, the single crystal XRD has been used to confirm the structure and the orientation of the single crystal. The results revealed the lattice parameters of the specimen are a = b =4.22\r{A}, c = 14.47\r{A}, $\alpha$=$\beta$=$\gamma$=90$^\circ$ with the tetragonal structure at room temperature. The atomic proportion was determined by energy dispersive x-ray spectroscopy (EDS). The sample for measurements with the dimension of $\sim 2\times$ 0.5 $\times$0.1 $mm^{3}$ and the magnetic field was aligned along $c$-axis in all measurements. The low-field magnetic transport measurements were performed on an Integra AC (Oxford Instruments) with a 16 T superconducting magnet and a Leiden dilution refrigerator with a 14 T superconducting magnet. Temperature- and field-dependent resistivity measurements were made in the standard four-probe method with a pair of current source (Keithley 6221) and DC-Nanovoltmeter(Keithley 2182A). The high-field magnetic transport measurements were carried out under pulsed magnetic field in Wuhan National High Magnetic Field Center (WHMFC). Golden wires were attached using silver paste on the rectangular sample and the every contact resistance were maintained to be less than 2 $\Omega$ in the measurements.

\begin{figure}[!t]
	\includegraphics[width=8.5cm]{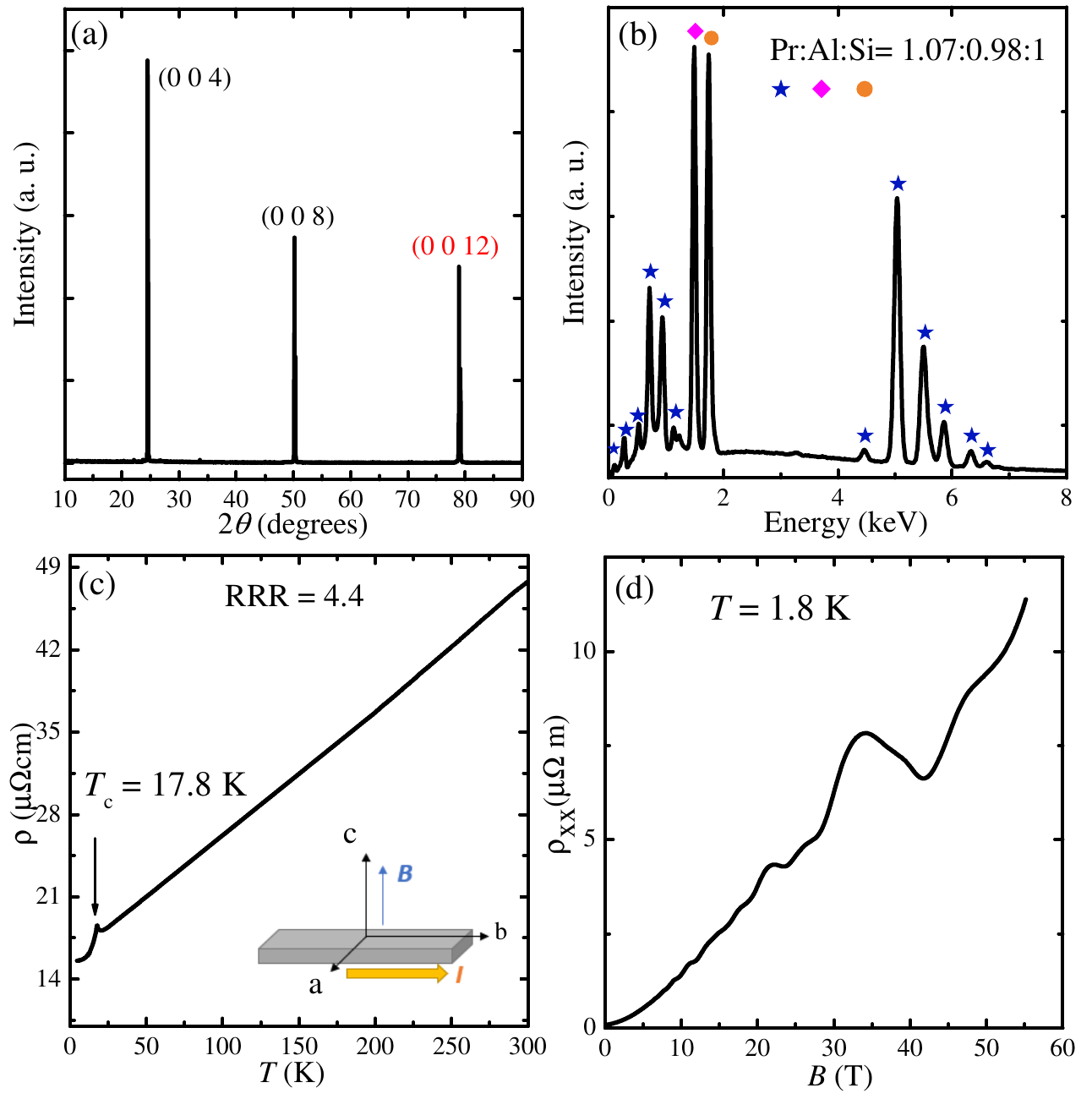}
	
	\caption{(a) X-ray diffraction pattern of PrAlSi single crystal. (b) The diagram of the energy dispersive x-ray spectroscopy (EDS).  (c) Temperature dependence of electrical resistivity from 2-300 K of the PrAlSi single crystal. The residual resistivity ratio $\rho$$ {(300 K)}$/$\rho$${(2K)}$ is about 4.4. Inset is the configuration of the measurements. (d) Longitudinal resistivity as a function of magnetic field up to 55 T at 1.8 K.
		\label{fig:sample}}
\end{figure}

The XRD pattern of the PrAlSi single crystal shown in Fig. \ref{fig:sample}(a) indicating the high quality sample in our measurement and the compositions of the synthesized crystals were analyzed using energy dispersive spectroscopy (EDS), showing in the Fig. \ref{fig:sample}(b). \textcolor{black}{The XRD data in our work  obtained from single crystal with selected plane (0 0 1) and in the previous studies (ref.[24] and ref.[43]) obtained from synthesized powder make the difference in the XRD patterns.} The temperature - dependent of longitudinal resistivity $\rho_0 {(T)}$ of the PrAlSi single crystal in the absence of an external magnetic field in this work was measured with the current applied along the $b$-axis ([010]), which showing a typical semimetal character, is presented in the Fig. \ref{fig:sample}(c). At high temperature, $\rho_0 {(T)}$ displays the metallic behavior with an almost linear temperature dependence from 300 K and develops a small cusp at ferromagnetic transition $T_{\rm{c}}$ = 17.8 K before further cooling to 2 K. The residual resistivity ratio RRR which is defined as $\rho {(300 K)}$/$\rho {(2K)}$ reaches the value of approximate 4.4 in our case. Fig. \ref{fig:sample}(d) displays the longitudinal magnetoresistivity up to ~1140$\mu\Omega$cm under high magnetic field.

\section{Raw data of the field dependence of magnetoresistance and average mobility}
\begin{figure}[!t]
	\includegraphics[width=8.5cm]{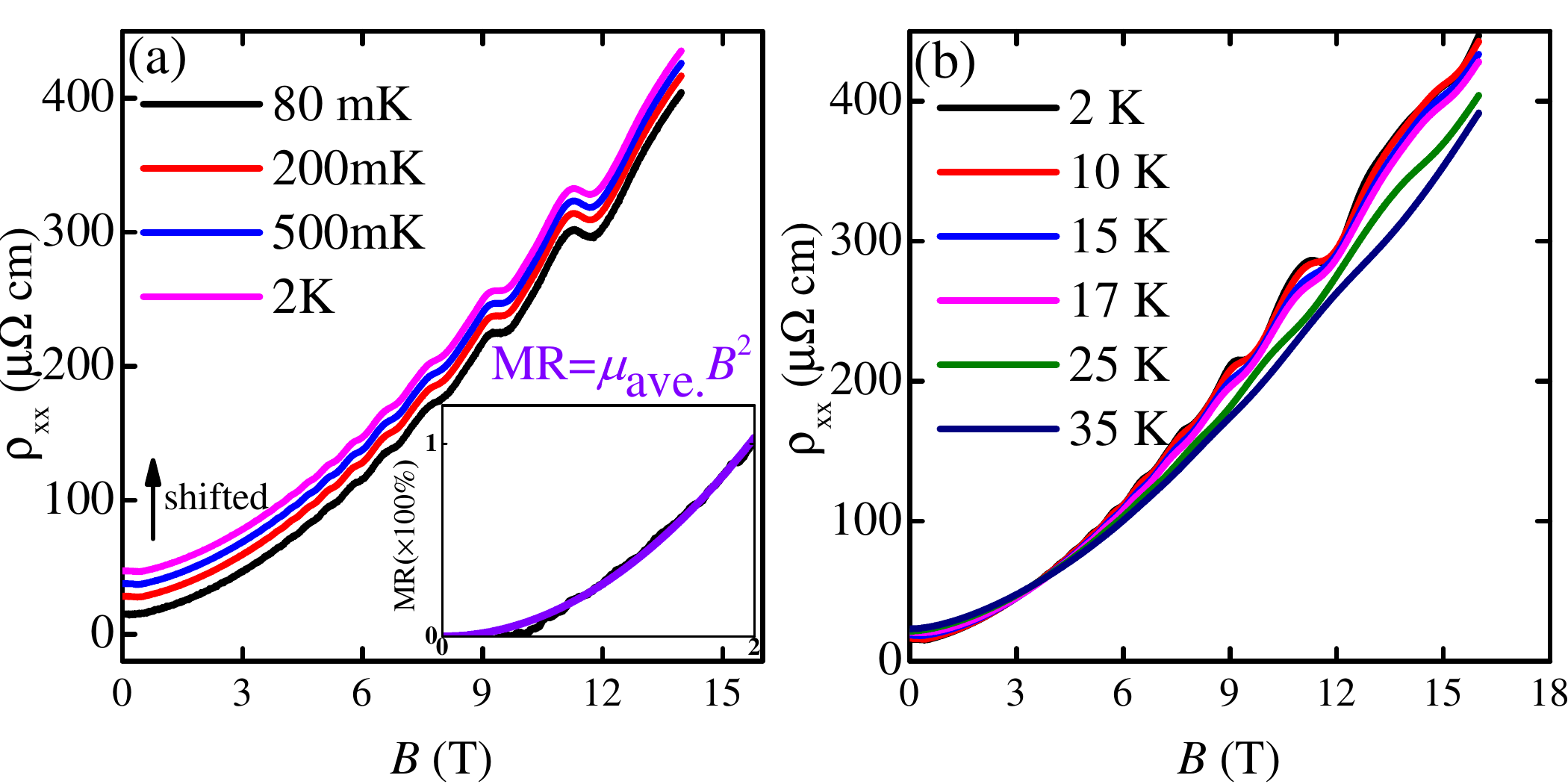}
	
	\caption{(a) The temperature-dependent magnetoresistivity $\rho_{xx}$ (shifted) measured from $T$ = 80 mK to 2 K with magnetic field up to 14 T, the inset is the fitting result of the equation MR=$\mu_{ave}B^2$ with $\mu_{ave}$= 0.26 m$^2$/Vs. (b) $\rho_{xx}$ with the temperature variation from 2 K to 35 K as magnetic field up to 16 T.}
\label{fig:MR}
\end{figure}

The temperature dependence of magnetoresistance of single crystal PrAlSi were measured under static field up to 14 T (Leiden dilution refrigerator) in Fig. \ref{fig:MR}(a) and up to 16 T (Oxford Instruments) in Fig. \ref{fig:MR}(b), respectively. Positive and unsaturated magnetoresistance and evident SdH oscillations under low temperature above 3 T are visible until up to 35 K in our measurements. The inset in (a) is the fitting result of MR=$\mu_{ave}B^2$ guided by the violet line and the obtained average mobility is 0.26 m$^2$/Vs.

\section{Details of FFT analyses}

\begin{figure}
	\includegraphics[width=8.5cm]{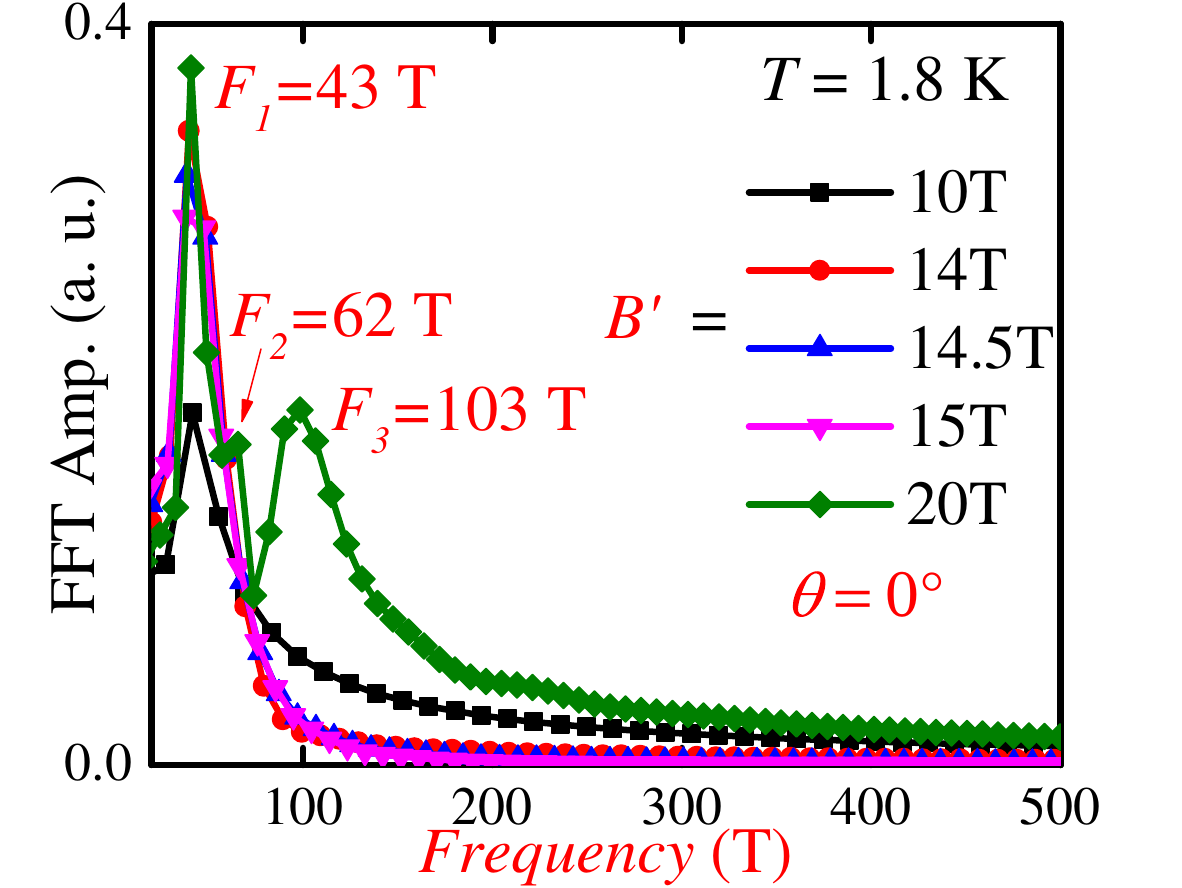}
	
	\caption {The FFT analysis spectrum of MR \textcolor{black}{in the reciprocal fields from 5.7 T to the selected magnetic field $B'$ near $B^*$,} measured in pulsed field.}
\label{fig:FFT analysis with slected B}	
	
\end{figure}

\begin{figure}
	\includegraphics[width=9cm]{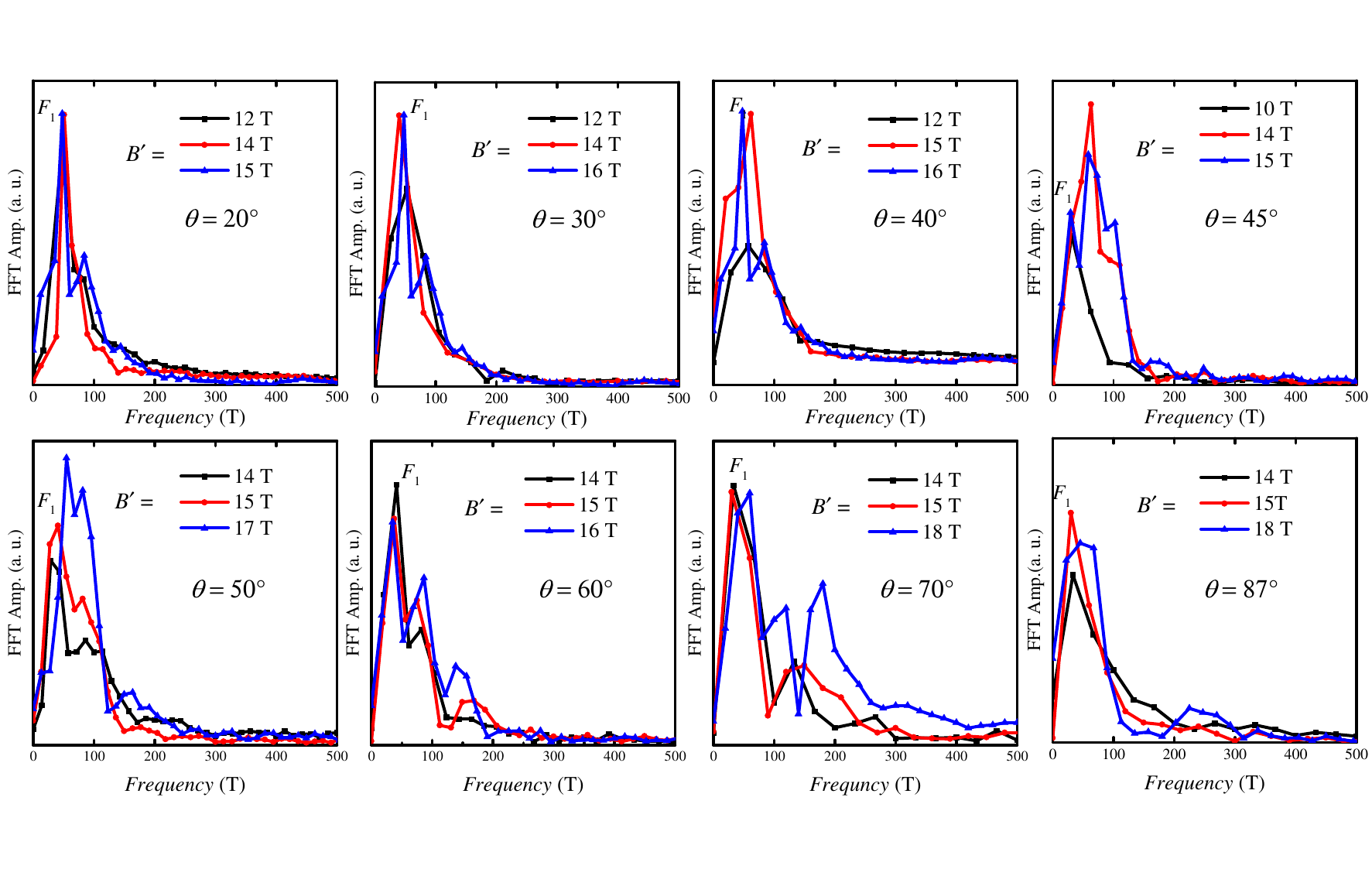}
	
	\caption{The FFT analysis spectrum of MR in the reciprocal fields from 8 T to the selected magnetic field $B'$ at different angles.}
\label{fig:FFT analysis with angles}	
\end{figure}

As mentioned above, the SdH QOs frequencies are discriminative with only one dominating frequency $F{_1}$ (43 T) below 14.5 T and another two higher frequencies $F{_2}$(62 T) and $F{_3}$(103 T) exist under higher field. We selected different low magnetic field ranges to analyze the FFT from the result of MR \textcolor{black}{with {$\theta$} = 0$^{\circ}$} showing in Fig. \ref{fig:FFT analysis with slected B}. The FFT results accord with the discussion in Fig. 2, which indicating only one domain frequency under the critical magnetic field 14.5 T and the FFT peak becomes widely at 15 T and involves into another peaks when magnetic field increases up to 20 T eventually. This FFT analysis clearly shows the frequency variation as the function of magnetic filed, we consider that 14.5 T (labelled with $B^{\ast}$) is indeed a critical point and the conclusion is well agreed with the two-band fitting of Hall resistivity in main text. \textcolor{black}{ In addition, in Fig. \ref{fig:FFT analysis with angles}}, similar to the method as mentioned above, the segmented FFT \textcolor{black}{analyses} with different angles (from 20$^{\circ}$ to 87$^{\circ}$) were carried out and the results were summarized in the inset of Fig. 5(b).

\section{The Lifshitz-Kosevich (LK) theory}

As is known, the oscillatory part of magnetoresistance can be described by the LK formula\cite{shoenberg2009} which is:
$\Delta\rho=\sum_{i} A_{i}R_{Ti}R_{Di}{\rm {cos}}[2\pi(\frac{F_i}{B}-\delta_i)] $, where $i$ is the number of Fermi pocket, $A_{i}$ are prefactors, $F_i$ are the oscillatory frequencies and  $\delta_i$ are the phase factors. The $R_{Ti} = \alpha Tm^{\ast}_i/B{\rm {sinh}}[\alpha Tm^{\ast}_i/B]$ and $R_{Di}={\rm {exp}}[\alpha T_{Di}m^{\ast}_i/B]$ are the thermal and scattered damping factors, where $\alpha=2\pi^2k_Bm_e/e\hbar\simeq14.69$ T/K, $m^{\ast}$ is the cyclotron mass and $T_{Di}$ is the Dingle temperature.

\section{Field dependence of mass for the $F_1$}
\begin{figure}[t]
	\includegraphics[width=8.5cm]{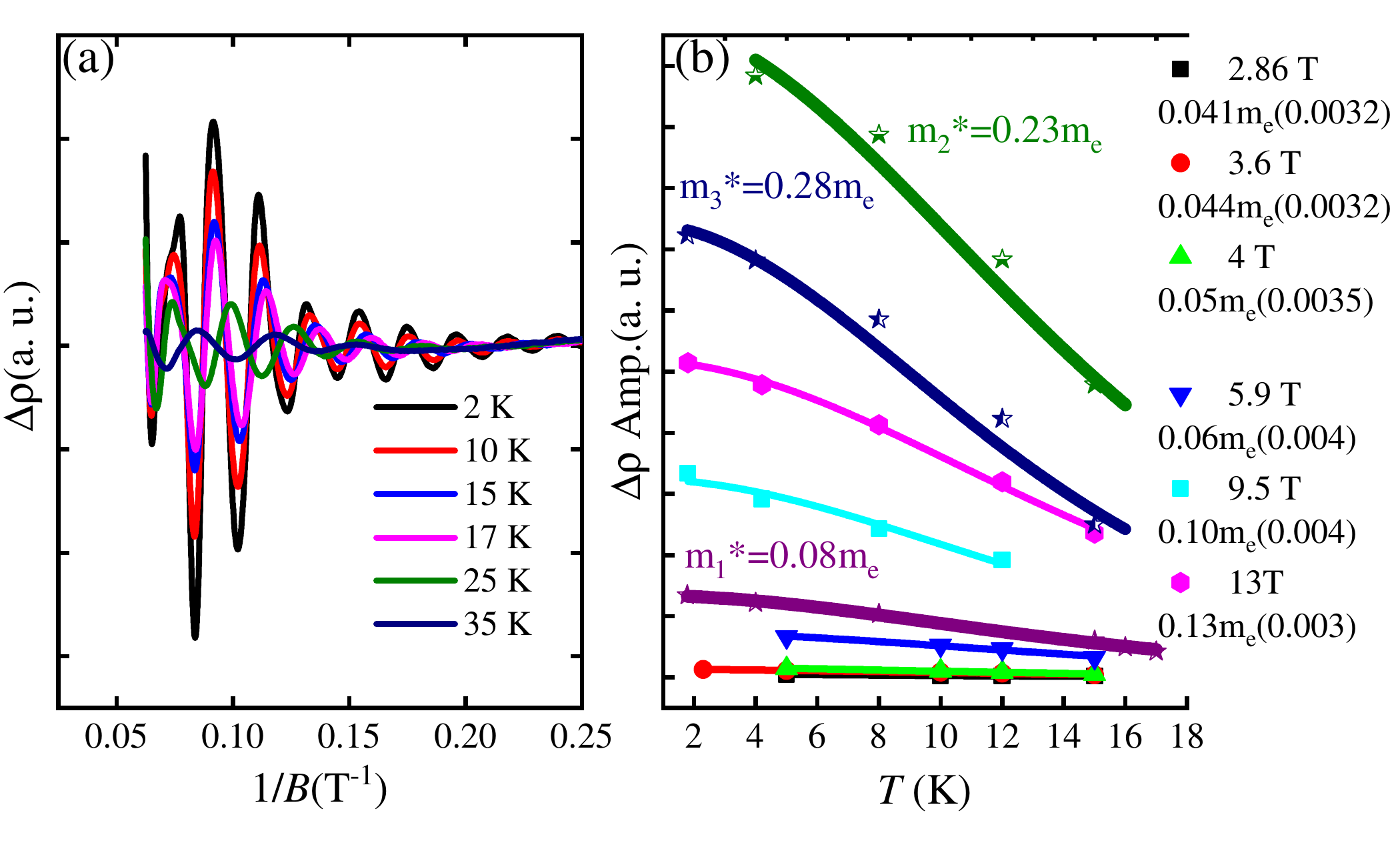}
	
	\caption{(a)The SdH oscillatory component as a function of 1/$B$ after subtracting the smoothed background from 2 K to 35 K measured under static field with magnetic field up to 16 T. (b) Effective masses of $F_1$ as a function of magnetic field below $B^{\ast} $ extracted from the temperature-dependent SdH amplitude of peaks by fitting with the damping factor in the LK function are summarized in the Fig. 2(c). Besides, the cyclotron mass $m^*$ of $F_1$ was extracted by fitting the FFT amplitude of SdH oscillations (Fig. 2(a)) to the temperature damping factor of the LK equation from 5.7 T to 14.5 T, and the same method for $F_2$ and $F_3$ with field from 14.5 T to 55 T. }
\label{fig:effective mass}
	
\end{figure}

As described in main text, the LT occurs around $B^{\ast}=$14.5 T with the disappearance of a lower SdH frequency $F_1$ = 43 T. It's naturally to study the evolution of this Fermi pocket so that we extracted the effective mass as a function of magnetic field below $B^{\ast}$ from the temperature dependence of the SdH oscillatory components at different fields as shown in the Figure. \ref{fig:effective mass}. We can extract m$^*$ from the SdH peaks directly because it relates to the temperature damping factor which is evident reflecting in the SdH oscillation component showing in Figure. \ref{fig:effective mass} (a). In addition, the effective masses of the Fermi pockets mentioned in Fig. 2, which were extracted from the amplitude of temperature-dependent FFT spectra, are plotted in Figure. \ref{fig:effective mass} (b).

\section{Quantum oscillation Phase factor analysis for $F_1$}

\begin{figure}[!t]
	\includegraphics[width=8.5cm]{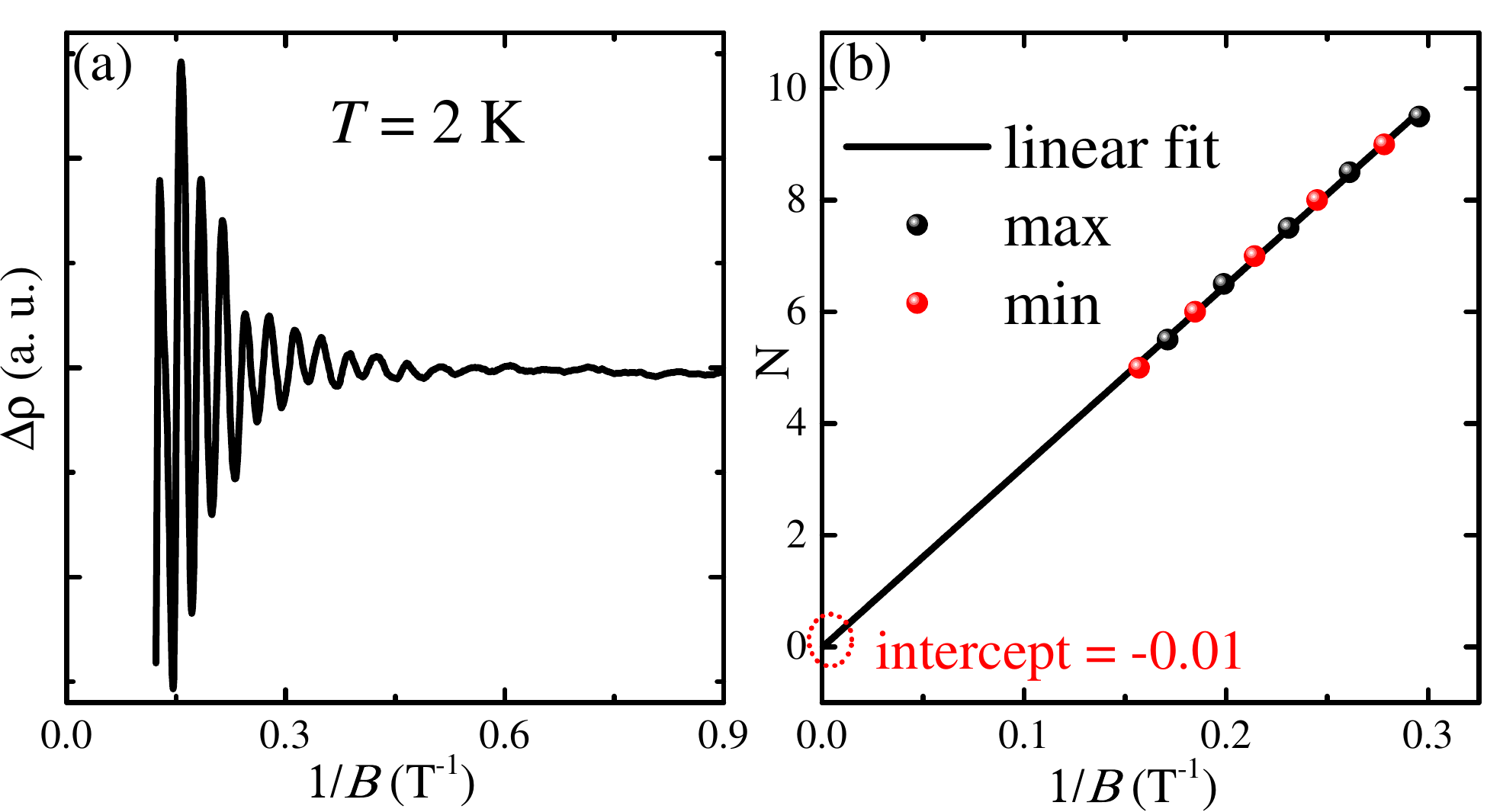}
	
	\caption{(a) $\Delta{\rho}$ as the function of $B^{-1}$ at indicated temperature.  (b) Landau level fan diagram of $F{_1}$ extracted from $\Delta{\rho}$ and the line indicates the linear fitting to the data. The solid circles represent the maxima and minima of $\Delta{\rho}$, respectively.}
	
\label{fig:LL}
\end{figure}

In order to pin down the property of the topological state in PrAlSi, the Landau level (LL) index fan diagram, which could manifest the nontrivial $\varphi{_B}$ in Weyl system, is much essential. The oscillatory component $\Delta{\rho}$ at 2 K under static field is obtained by subtracting the background and plotted as the function of 1/$B$ in Figure. \ref{fig:LL} (a). An apparent simple pattern here is in accordance with the discussion of Fig. 1 in main text indicating the single main frequency under low magnetic field range in QOs pattern.
According to the LK formula:

\begin{equation}
	\Delta{\rho}\sim{\frac{\lambda{T}}{\rm{sinh(\lambda{T})}}e^{-\lambda{T{_D}}}cos[2\pi(\frac{F}{\rm{H}}-\frac{1}{\rm{2}}+\delta)]}
\end{equation}

Here, the phase factor is $\delta-\frac{1}{\rm{2}} = \frac{\varphi{_B}+\varphi{_D}}{\rm{2\pi}} -\frac{1}{\rm{2}}$, where $\varphi{_B}$ is Berry phase and $\varphi{_D}$ equals $\pm\frac{1}{8}$ which depends on the cross-section extremum with maximal or minimal for 3D cases. Figure. \ref{fig:LL} (b) displays the Landau level(LL) fan diagram for the fundamental frequency $F{_1}$. Here, the minimum $\Delta{\rho}$ (red) is integer indices and the maximum $\Delta{\rho}$ (black) is half-integer indices. All points are well linearly fitting and we extrapolate $1/B$ to zero and obtain the intercept is -0.01 =  $\delta-\frac{1}{\rm{2}}$, indicating a $\pi$ Berry phase and non-trivial topological state of the $F{_1}$ in PrAlSi single crystal.

\section{Two-band fitting for the magnetoresistivity $\rho_{xx}$ and Hall conductivity $\sigma_{xy}$}

$\rho_{xx}$ and $\sigma_{xy}$ in our case can be described by the two-band model:

\begin{equation}
	\rho{_{xx}}=\frac{1}{\rm{e}}\frac{(n{_h}\mu{_h}+ n{_e}\mu{_e})+\mu{_h}\mu{_e}(n{_e}\mu{_h}+n{_h}\mu{_e})B^2}{(n{_h}\mu{_h}+ n{_e}\mu{_e})^2+\mu{_h}^{2}\mu{_e}^{2}(n{_h}-n{_e})^2B^2}
\end{equation}
\begin{equation}
	\sigma{_{xy}}=eB\left(\frac{n{_h}\mu{_h}^{2}}{1+\mu{_h}^2B^{2}}-\frac{n{_e}\mu{_e}^{2}}{1+\mu{_h}^2B^{2}}\right)
\end{equation}

\begin{figure}[b]
	\includegraphics[width=8.5cm]{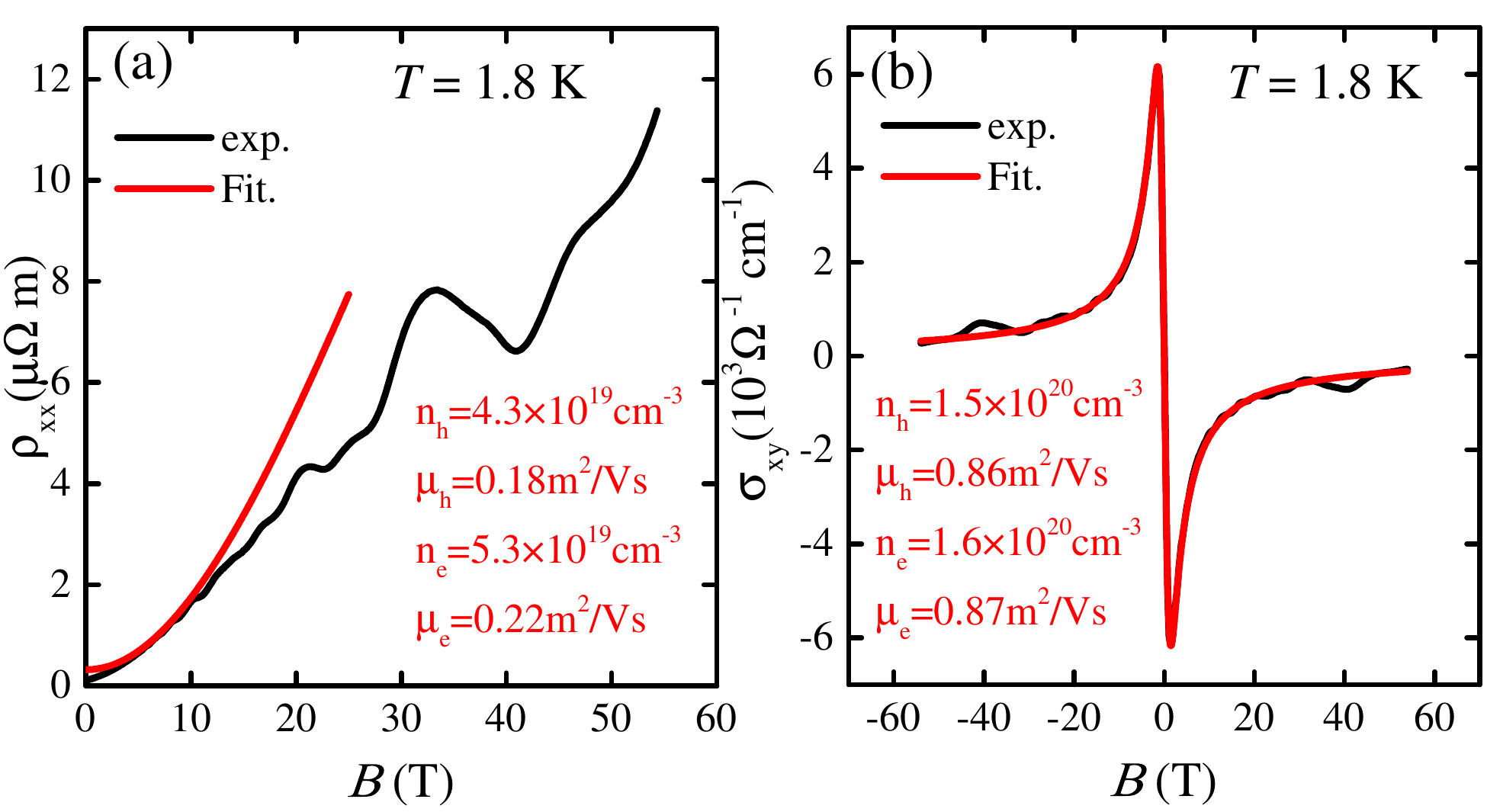}
	
	\caption{(a) and Hall conductivity (b) of PrAlSi at 1.8 K. The black solid line is the experimental data and the red solid line is the fitted curve with two-band model, respectively.}
\label{fig:Hall conducivity}
\end{figure}

\begin{figure}
	\includegraphics[width=8.5cm]{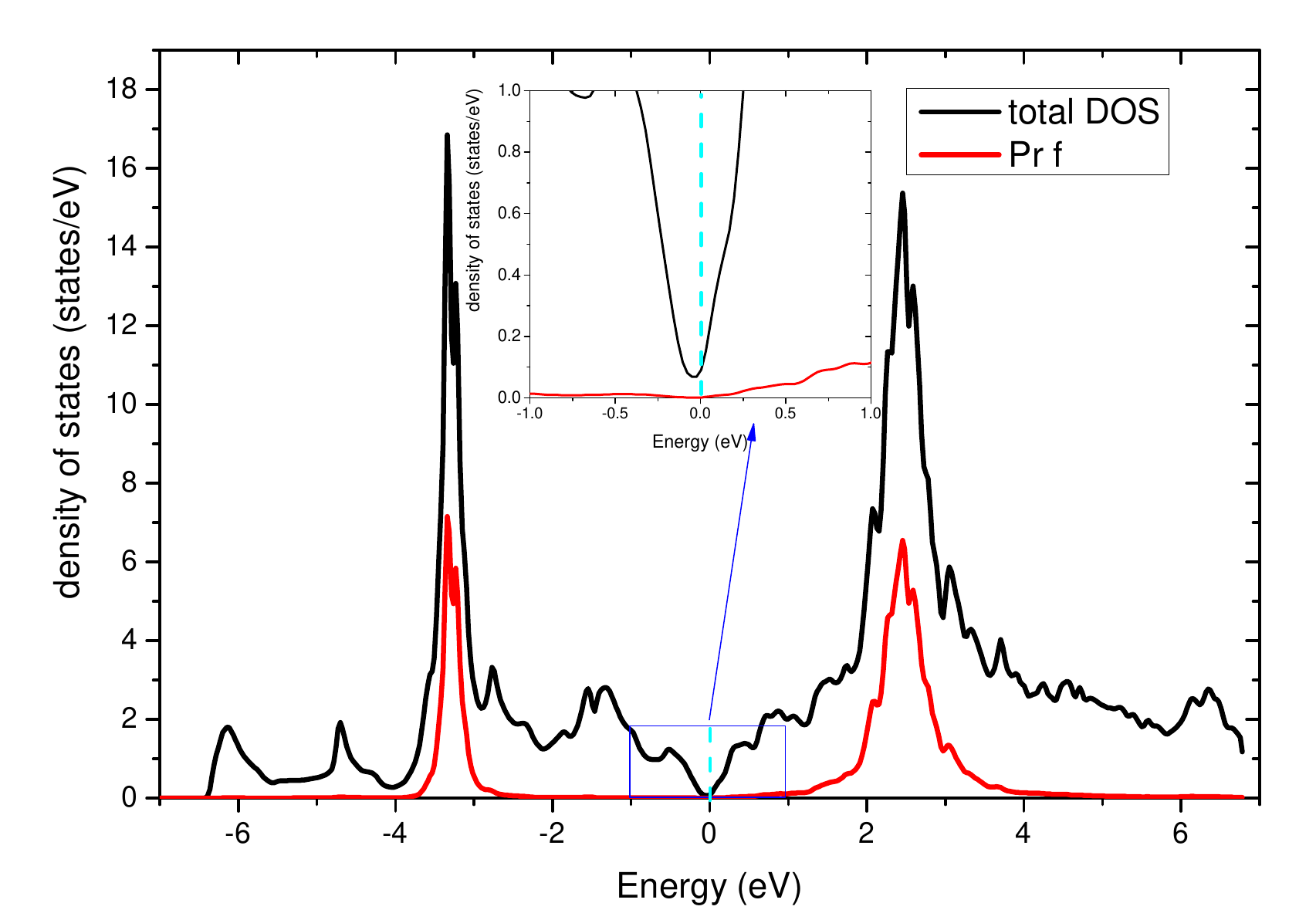}
	
	\caption{The total density of states and the density of states (DOS) of $f$ orbit.}
	
\label{fig:DOS}	
\end{figure}

Here, $n{_h}(n{_e})$ and $\mu{_h} (\mu{_e}$) are carrier concentration and mobility for the holes (electrons), respectively.  The fitted result of magnetoresistivity $\rho_{xx}$ with the parameters identical with the main text shows a good fit in low field range up to $B^{\ast}$ shown in Figure. \ref{fig:Hall conducivity} (a) in red, while the experimental curve is in black. 
The fitted result of Hall conductivity $\sigma_{xy}$ indicates carrier concentration are $n{_h} =1.5\times10^{20}cm^{-3}$ ($n{_e}=1.6\times10^{20}cm^{-3}$) and the corresponding mobilities are $\mu{_h}$= 0.86 m$^{2}$/Vs($\mu{_e}= 0.81m^{2}$/Vs). The field dependence of Hall conductivity shows a sharp rise in the low fields and slowly decrease as the field is increased. The fitting is tricky and a little change in the parameters would not change the fitting result, especially in the high-field range. So we fit the Hall resistance which has much better shape for the fitting, giving more convincing results. Besides, the obtained carriers 1.5$\times$10$^{20}$ (1.6$\times$10$^{20}$) cm$^{-3}$ for holes (electrons) lead to a mobility of 0.17 m$^2$/Vs at zero field, which is quite distinct from the fitting results of $\sim$0.9 m$^2$/Vs.

\section{The comparison of Hall resistivity with CeBiPt}

In CeBiPt, LT promotes carrier density from 7.2$\times$ 10$^{17}$ cm$^{-3}$ to 9.2$\times$ 10$^{17}$ cm$^{-3}$, increased by 28\%. The change of PrAlSi is mainly in mobility beyond after LT, with 75\% for holes and 27\% for electrons. The large drop of mobility is most likely due to the increase of inter-bands scattering because of more pockets present after LT and also the increase of mass of the pockets. The drops in mobility may result in the downward of MR above LT in the Fig. 1(a). The sub-linear Hall resistance is single-band behavior in CeBiPt\cite{Kozlova2005}, while it is a multi-band behavior in PrAlSi and has two-order higher carriers concentration and mainly changes its mobility. Thus, one can clearly observe the linearity change of Hall resistance in CeBiPt, while smooth evolution of Hall resistivity and an evident deviation in first derivative of Hall resistivity in the current case. The same feature is that no sudden change occurs in Hall resistance  during LT. 

\begin{figure}
	\includegraphics[width=8.5cm]{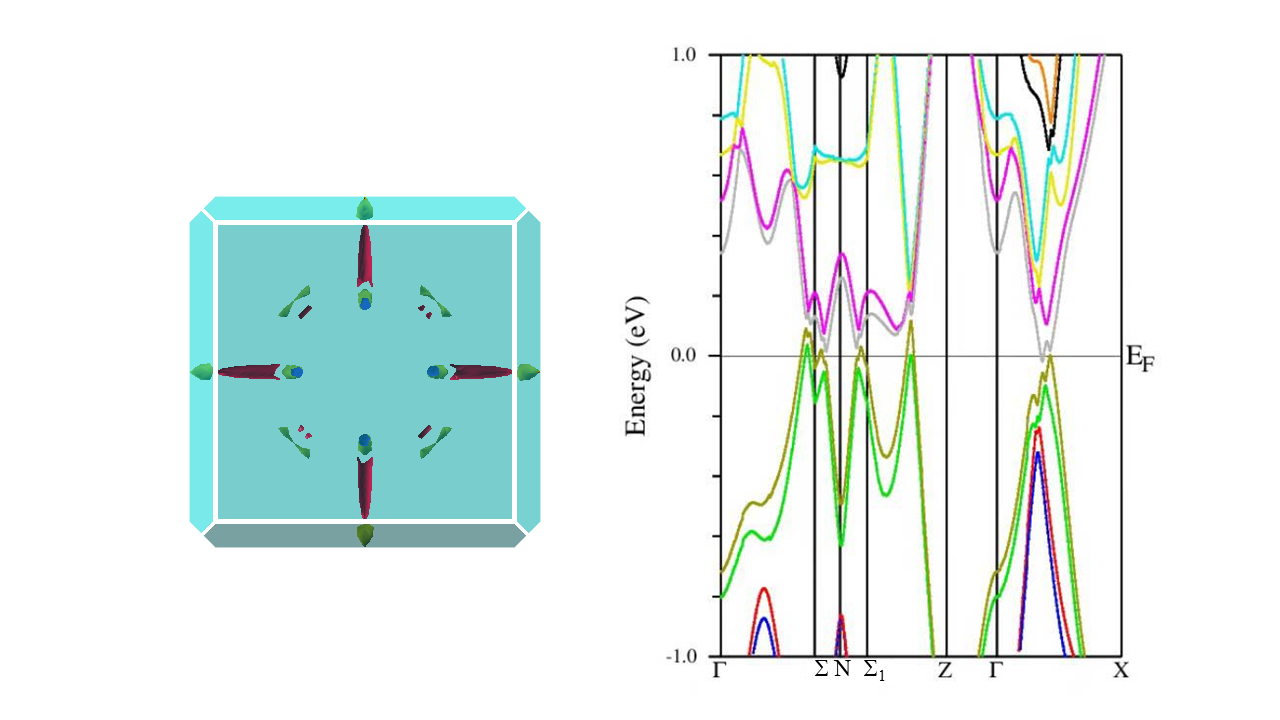}
	
	\caption{The first-principle calculation results. (a) The calculated Fermi surface and (b) band structure of the PrAlSi.}
\label{fig:calculation}	
\end{figure}

\section{Internal field and external field}

PrAlSi is a ferromagnet with the $T{_c}=17.8$ K (Fig. S1(c)). We considered its demagnetization factor and external field to infer the internal field which exerts on carriers. The internal field $B_{int} = \mu_{0}H_{ext}+ \mu_{0}(1-n)M$, where $n$ is the demagnetization factor. According to the report \cite{aharoni1998}, n = 0.77 of our rectangular sample is obtained by fitting the shape factor into the given function. $\Delta B$ =$ B_{int}-B_{ext}$=0.23$\mu_{0}M_{exp}=$0.46 T, here $\mu{_0}$=$1.257\times10^{-6}H/m$.

\section{The DOS of Pr f electron}

The Fig. \ref{fig:DOS} shows the total density of states and the density of states (DOS) of $f$ orbit from Pr in the energy scale of -7 to 7 eV. And the inset shows the enlarged part for the energy scale of -1 to 1 eV. All the $f$ electrons of Pr are clearly below Fermi level and do not contribute to the conductivity on the Fermi surface. So we will not expect $f-c$ ($c$-conduction electrons) hybridization in the system.

\begin{figure}[b]
	\includegraphics[width=9cm]{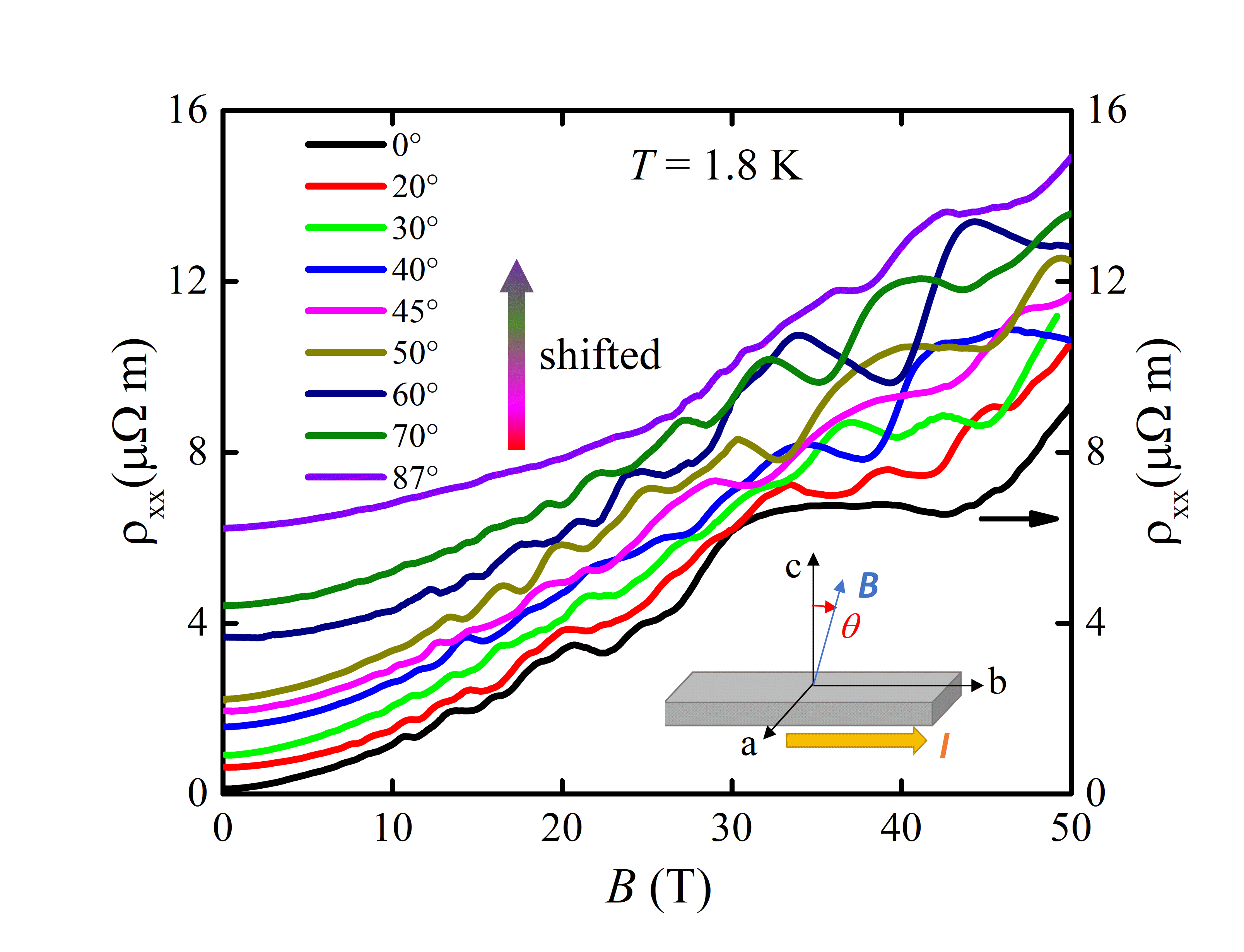}
	
	\caption{Isothermal magnetoresistance measured at different angles. The magnetic field dependence of the resistivity with {$\theta$} = 0$^{\circ}$ follows the right scale and the curves of other angles were shifted  as the arrow indicated. Inset is the configuration of the measurements.} 
\label{fig:raw angular}	
\end{figure}

\begin{figure}
	\includegraphics[width=8cm]{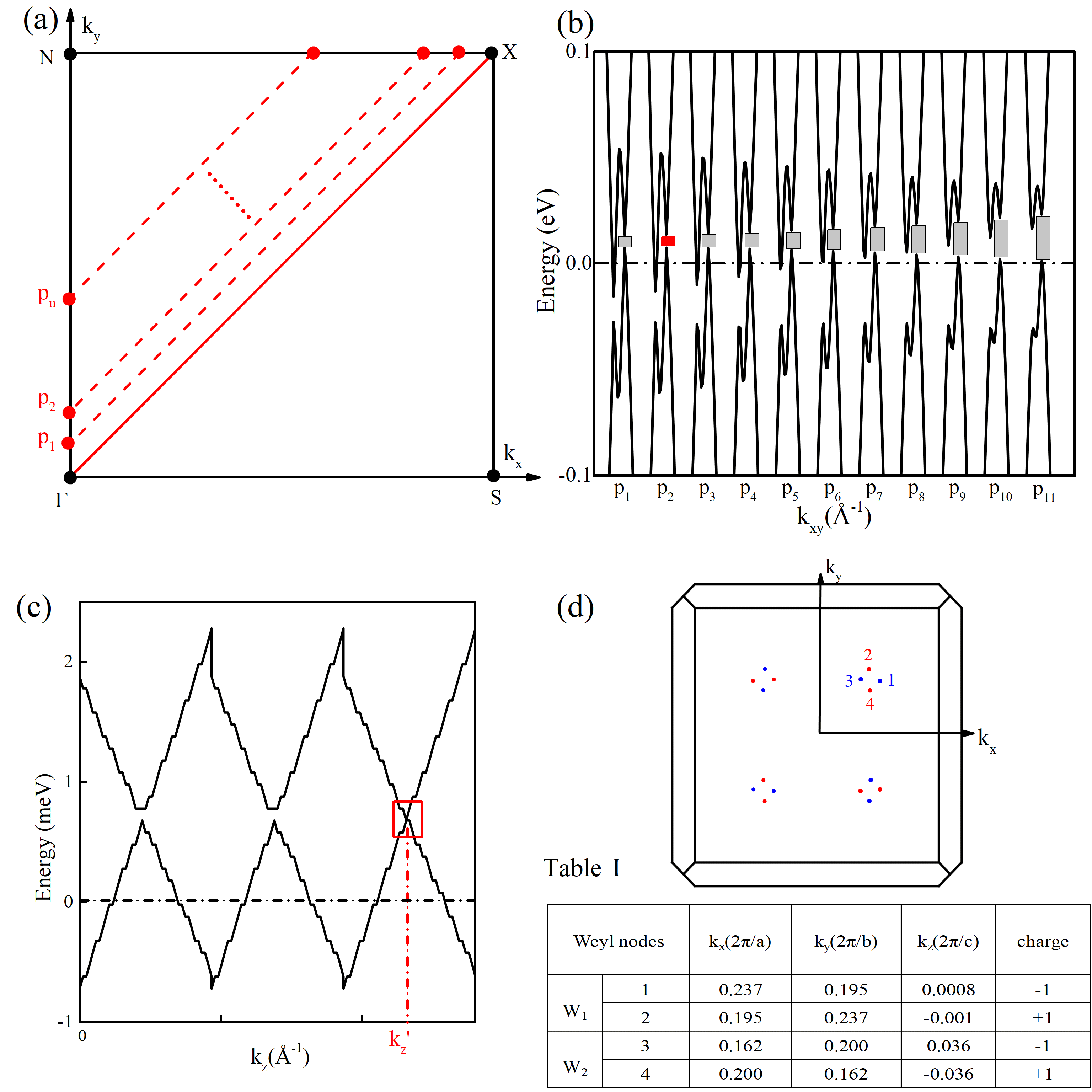}
	
	\caption{(a), (b) and (c) introduce the process of band calculation for obtaining the Weyl point 1 showing in Fig. \ref{fig:Weyl}(d). (a) A diagram introduces the selective paths which are parallel to the $\Gamma$-X direction ( in the plane with $k_z$ = 0) in the calculation in detail. (b) The results of the calculated band structure according to the paths showing in graph (a) with different values of $k_x$,$k_y$ and the rectangle in red indicates the projected position of the Weyl point. (c) Further calculation with only changing $k_z$ for getting the gapless point(the exact value of $k'_z$) labeled by the red rectangle. (d) The four key Weyl points are illustrated in the sketch. Table \uppercase\expandafter{\romannumeral1} presents the specific position of nodes 1-4. }
	
\label{fig:Weyl}	
\end{figure}

\section{Theoretical calculation}
The theoretical calculation was carried out with WIEN2k which is based on the density functional theory (DFT) based linearized augmented plane wave (LAPW) method\cite{Blaha2020}. We used the Perdew-BurkeErnzerhof (PBE) generalized gradient approximation (GGA)\cite{Perdew1996} form of the exchange correlation function. The RKmax was set to 5 which is determined by the smallest atom Si\cite{Blaha2020}. The results are for the effective Hubbard U of 6 eV by taking the on-site correlation using the DFT+U approximation for the $f$ state of Pr. The spin-orbit coupling was included in “second variation” of the calculation. The calculated results are consistent with the reported\cite{yang2020PrAlSi}, shown in the Fig. \ref{fig:calculation}, plotted by XcrysDen\cite{Kokalj1999}. Along $\Gamma$-X direction in the Fig. \ref{fig:calculation} (a), the small pockets located on the sides of the direction have 8 pockets, which could be the $F_1$. The calculated band structure shown in the Fig. \ref{fig:calculation} (b) is also well consistent with the reported. Along $\Gamma$-X direction, the bands are shallow and could be source of the instability during the LT process. In Fig.5(d) of the main text, the data at the angles larger than 60$^{\circ}$ are the average values because of the data calculated from the SKEAF program were noisy, which probably due to the insufficient $k$-points in the calculation.

\section{Raw data of MR at different tilt angles $\theta$}

The angle dependence of quantum oscillations at different angles would provide much information of Fermi surface, such as shape and dimensionality. Fig. \ref{fig:raw angular} shows magnetoresistance at different $\theta$ values with pulsed field at 1.8 K. A shifted figure was shown to clarify the MR with tilted angles. As seen, all the magnetoresistance are positive with the increasing angles and the oscillations are evident, which indicating a 3D feature of the Fermi surface.

\section{Weyl points ascertaining by band structure calculation}

The first step in our case is to address the points in $k_x$-$k_y$ plane with $k_z$ = 0, where the tiniest gap band occurs.Fig. \ref{fig:Weyl} (a) is a rough schematic diagram with the paths of the band calculation. The red solid and dotted lines are the equally spaced paths in the direction parallel to the $\Gamma$-X line of the BZ. The results of calculated band structure are shown in Fig. \ref{fig:Weyl} (b) with red rectangular frame indicating the approximate path, where the Weyl points locate. And then, as Fig. \ref{fig:Weyl} (c) shown, changing the values of $k_z$ for getting the gapless points to ensure accurate position of the Weyl points. Clearly, the Weyl point 1 is of hole nature.  Fig. \ref{fig:Weyl} (d) shows the partial Weyl points in the first Brillouin zone and the corresponding positions in $k$-space of the points (1-4) are listed in Table \uppercase\expandafter{\romannumeral1}.

\end{document}